\documentclass[12pt]{article}
\pdfoutput=1
\usepackage{jheppub}
\usepackage{amsfonts}
\usepackage[section]{placeins}
\usepackage{amsmath,amssymb} 
\usepackage{graphicx} 
\usepackage[dvipsnames,x11names]{xcolor} 

\usepackage[nottoc,notlof,notlot]{tocbibind} 
\usepackage[titles]{tocloft} 

\usepackage{float}
\usepackage{subcaption}
\usepackage{tikz}
\numberwithin{equation}{section} 

\title{\textbf{Thermalization in different phases of charged SYK model}}

\author[a]{Tousik Samui}\emailAdd{tousiksamui@hri.res.in}

\author[b]{and Nilakash Sorokhaibam}\emailAdd{nilakash@niser.ac.in}

\affiliation[a]{
  Harish-Chandra Research Institute, 
  Homi Bhabha National Institute\\
    Chhatnag Road, Jhunsi, 
    Allahabad 211019, India}
    
\affiliation[b]{
National Institute of Science Education and Research, HBNI, Bhubaneswar 752050, Odisha, India}

  \abstract{We study thermalization of charged SYK model in two different phases. We show that both the highly chaotic liquid phase and the dilute gas phase thermalize. Surprisingly the dilute gas state thermalizes instantaneously. We argue that this phenomenon arises because the system in this phase consists of only long-lived quasi-particles at very low density. The liquid state thermalizes exponentially fast. We also show that the additional introduction of random mass deformation (q=2 SYK term) slows down thermalization but the system thermalizes exponentially fast. This is observed despite the fact that the addition of large q=2 SYK interaction forces spectral statistics to obey Poisson statistics. An interesting new observation is that the effective temperature is non-monotonic during thermalization in the liquid state. It has a bump at relatively long time before settling down to the final value. With non-zero chemical potential, the effective temperature oscillates noticeably before settling down to the final value.}

\begin{document}
\maketitle

\section{Introduction and Summary}
\label{sec:intro}
Non-equilibrium dynamics of interacting quantum systems has been a subject of great interest for a long time \cite{Polkovnikov:2010yn,Gogolin:2016hwy}. It is a subject of interest in various fields of physics, e.g. various aspects of condensed matter physics, heavy ion collisions, AdS/CFT correspondence and black hole dynamics, quantum cosmology, etc. It also has wide applications in engineering sciences (e.g. working principle of everyday semiconductor devices), biophysics (e.g. application in protein folding), etc. It is also the underlying principle for the impending rise of quantum computers in the next few decades \cite{Altman:2019vbv}.

In this paper we will examine non-equilibrium dynamics of chaotic quantum systems. We will be considering closed quantum systems. It is generally expected that these systems thermalize. We will be considering thermalization in the sense that, starting from a thermal state, after some finite perturbation the system comes back to thermal equilibrium.\footnote{We will not be considering thermalization of pure excited states.} The most interesting aspect of this work is that we find certain special states which thermalize instantaneously.

Our interest is focused on Sachdev-Ye-Kitaev (SYK) models \cite{Sachdev:1993,Kitaev:2015,Maldacena:2016hyu}. These are chaotic quantum systems consisting of N fermions with q-body all-to-all random interactions.
\begin{equation}
\sum_{i_1,i_2,...,i_q=1}^N j_{q,i_1,i_2,...,i_q}\Psi^{\dagger}_{i_1}\Psi^{\dagger}_{i_1}\Psi^{\dagger}_{i_2}...\Psi^{\dagger}_{i_{q/2}}\Psi_{i_{q/2+1}}\Psi_{i_{q/2+2}}...\Psi_{i_{q}}\
\end{equation}
The original SYK model consists of Majorana fermions but we will be considering complex fermions with which the model has a conserved charge other than the Hamiltonian.
\begin{equation}
{\bf Q}=\frac{1}{N}\sum_{i=1}^N \Psi^{\dagger}_{i}\Psi_{i}\
\end{equation}
For the rest of the paper, we will refer to a single interaction term as, for example, $q=2$ SYK term. This interaction term can be a part of a Hamiltonian consisting of multiple SYK interaction terms. We will refer to the full model as, for example, $(q=2,4)$ SYK model. This model has both $q=2$ and $q=4$ SYK terms. Another example is $(q=2,4,6)$ SYK model which will consists of $q=2$, $q=4$, and $q=6$ SYK interaction terms.

In \cite{Sorokhaibam_2020}, the different phases of charged SYK model have been studied in the presence of chemical potential. It has also been examined if $(q=2,4)$ SYK model undergoes a similar phase transition (in the absence of chemical potential). In this work, we will study non-equilibrium dynamics in different phases of charged SYK model and $(q=2,4)$ SYK model. We do this by taking the systems out of equilibrium by performing quantum quenches. We will solve the non-equilibrium Green's functions using their equations of motion. We will be using lesser Green's function $G^<(t_1,t_2)$, greater Green's function $G^>(t_1,t_2)$, retarded Green's function $G^R(t_1,t_2)$ and advanced Green's function $G^A(t_1,t_2)$. The definition of these different Green's functions are as follows
\begin{eqnarray}
\label{lesserG}
G^<(t_1,t_2)&=&i\langle \psi^{\dagger}(t_2)\psi(t_1) \rangle\\
\label{lesserG}
G^>(t_1,t_2)&=&-i\,\langle \psi(t_1)\psi^{\dagger}(t_2) \rangle\\
\label{GR}
G^R(t_1,t_2)&=&\Theta(t_1-t_2)\left[G^>(t_1,t_2)-G^<(t_1,t_2)\right]\\
\label{GA}
G^A(t_1,t_2)&=&\Theta(t_2-t_1)\left[G^<(t_1,t_2)-G^>(t_1,t_2)\right]\\
G^K(t_1,t_2)&=&\left[G^>(t_1,t_2)+G^<(t_1,t_2)\right]\
\end{eqnarray}
Non-equilibrium dynamics of SYK models and other related models have been studied in \cite{Eberlein:2017wah,Bhattacharya:2018fkq,Haldar:2019slc,Maldacena:2019ufo,Almheiri:2019jqq}. Pure excited states of SYK models have also been studied in \cite{Kourkoulou:2017zaj,Dhar:2018pii,Numasawa:2019gnl}.

We will briefly clarify on what we meant by a chaotic quantum system. There are various diagnostics of quantum chaos. The two most popular and well studied diagnostics are comparison of spectral statistics with random matrix theory (BGS conjecture, \cite{Bohigas:1983er}) and exponential decay of Out-of-Time-Ordered correlators (OTOC) \cite{Maldacena:2015waa}. For a system like SYK model with widely separate time scales of dissipation and scrambling, OTOC decays exponentially. Considering the operators to be the microscopic fermionic degrees of freedom, the OTOC is
\begin{eqnarray}
C(t_1,t_2)&=&\text{Tr}\langle e^{-\beta H_{SYK}/4}\Psi_i^{\dagger}(t_1)e^{-\beta H_{SYK}/4}\Psi_j^{\dagger}(0)e^{-\beta H_{SYK}/4}\Psi_i(t_2)e^{-\beta H_{SYK}/4}\Psi_j(0)\rangle\nonumber\\
&=& \mathcal{F}_0(t_1,t_2)-\frac{\mathcal{F}_1(t_1,t_2)}{N}\, e^{\lambda_L (t_1+t_2)/2}+\mathcal{O}(N^{-2})\
\label{four_pt}
\end{eqnarray}
where $H_{SYK}$ is the Hamiltonian. $\lambda_L$ is the Lyapunov exponent. We have taken the regularized OTOC as in \cite{Maldacena:2015waa}.

We will be working with two-body $q=2$, four-body $q=4$, and six-body $q=6$ all-to-all random interactions. So the general Hamiltonian is
\begin{eqnarray}
H_{SYK}(t)&=&\mu \sum_i\Psi^{\dagger}_i\Psi_i+\sum_{i,j=1}^{N}j_{2,ij}(t)\Psi_i^{\dagger}\Psi_j+\sum_{i,j,k,l=1}^{N}j_{4,ij;kl}\Psi_i^{\dagger}\Psi_j^{\dagger}\Psi_k\Psi_l\nonumber\\
&&\qquad \qquad +\sum_{i,j,k,l,m,n=1}^{N}j_{6,ijk;lmn}(t)\Psi_i^{\dagger}\Psi_j^{\dagger}\Psi_k^{\dagger}\Psi_l\Psi_m\Psi_n\
\label{hsyk}
\end{eqnarray}
Since we will be performing quantum quenches using $q=2$ and $q=6$ SYK terms, we have considered these interactions to be time-dependent.
The mass term introduces an effective chemical potential $\eta=\mu$.
We can also consider thermal states with explicitly chemical potential $\eta$ turned on in which case the total effective chemical potential is $\eta+\mu$.

\begin{figure}[h]
\begin{center}
\includegraphics[width=0.6\textwidth]{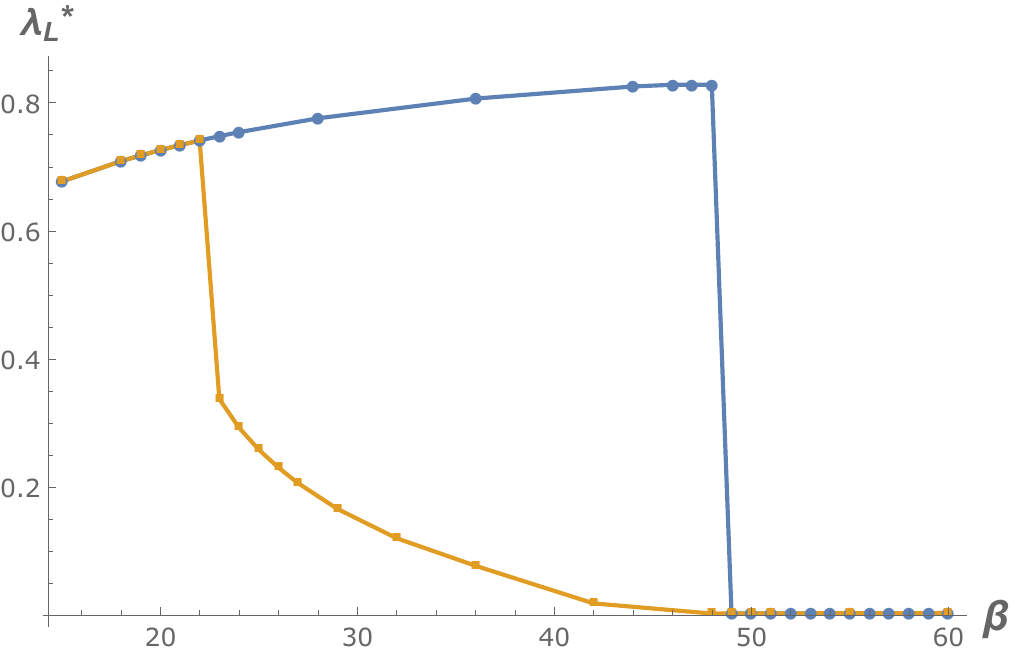}
\caption{\small Normalized Lyapunov exponent with varying inverse temperature in the two different phases for $\eta=0.27$ (reprinted from \cite{Sorokhaibam_2020}). The sharp changes in the blue (orange) curve represents the transition from the highly chaotic liquid (dilute gas) state to the dilute gas (liquid) state. So, the blue (orange) curve mostly constitutes the liquid (dilute gas) phase.}
\label{fig:phaselyap}
\end{center}
\end{figure}

It has been shown that in the presence of chemical potential there is a first order liquid-gas phase transition \cite{Choudhury:2017tax,Azeyanagi:2017drg,Sorokhaibam_2020}. The phase transition is between the highly chaotic state of the SYK model which we will call the liquid phase and the dilute gas phase.\footnote{The two phases are called chaotic phase and integrable phase in \cite{Sorokhaibam_2020}. We thank the anonymous referee for pointing out that it is most appropriate to call these phases as liquid phase and dilute gas phase.} Unlike the liquid phase, the dilute gas phase has a unique well-separated ground state so the zero temperature ($T\to 0$) entropy is $\log 1 =0$. This phase is also highly compressible as the name suggests. Again, as the name suggests, the charge or occupation number is also very low in this phase. For example,
\begin{equation}
\langle{\bf Q}\rangle=0.00035 \quad \text{at} \quad \beta=30 \quad \text{and}\quad \mu=0.27\
\end{equation}
while the occupation number for the free theory is $0.00030$. The phase diagram is produced in \cite{Azeyanagi:2017drg}. The phase transition happens in a finite range of the chemical potential. There is also a range of temperature in which the system can be in either of the two phases. So, one of the phases is metastable in this temperature range. In the liquid phase, the system is strongly interacting and there is no quasi-particle dynamics. While in the dilute gas phase, the system consists of long-lived quasi-particles at very low density. This is evident from the plot of the spectral function in the dilute gas phase. The spectral function is defined as the imaginary part of the retarded Green's function
\begin{equation}
A(\omega)=-2\,\text{Im}\, G_R(\omega)\
\label{spec_func_def}
\end{equation}
For the sake of simplicity the phase transition is always studied without the two-fermion random interaction $q=2$ term. The Lyapunov exponent of the liquid phase is suppressed exponentially when the chemical potential is turned on \cite{Bhattacharya:2017vaz,Sorokhaibam_2020}. This implies that chaos is suppressed. The Lyapunov exponent in the dilute gas phase has also been calculated in \cite{Sorokhaibam_2020}. It is very small at very low temperature. But it is large at relatively high temperature especially in the temperature range where the system can exist in either of the two phases. Figure \ref{fig:phaselyap} is a reprint of the plot of the normalized Lyapunov exponent for the two different phases with varying temperature and a fixed chemical potential.

It is conventional wisdom that quantum chaos implies thermalization. So the behaviour of the Lyapunov exponent suggests that turning on the chemical potential will slow down the thermalization process in the liquid phase. We will indeed find this to be the case. With the same reasoning, we expect that the dilute gas state would thermalize very slowly (or not at all) at very low temperature. At higher temperature in the dilute gas phase, we still expect that the system would still thermalize very slowly because the system still consists of long-lived quasi-particles at very low density. Surprisingly, we will find that the system in the dilute gas phase thermalizes instantaneously. The final state is a thermal state with a different temperature and a different chemical potential as compared to the initial state. We also like to point out that in purely (q=2) SYK model, the system equilibrates instantaneously but the final state is not a thermal state \cite{Bhattacharya:2018fkq}.

Highly chaotic systems thermalize exponentially fast \cite{Eberlein:2017wah}. So, we believe that the phenomenon of instantaneous thermalization of the dilute gas state has an underlying physics different from the chaotic dynamics. It is interesting that a simple harmonic oscillator (SHO) thermalizes instantaneously \cite{Sorokhaibam_2020}. This is true for both fermionic and bosonic SHOs. Consider the fermionic SHO Hamiltonian
\begin{equation}
H_{fSHO}=\mu \Psi^{\dagger}\Psi=\mu {\bf Q}
\end{equation}
Consider we start from a thermal state of this system with the density matrix $e^{-\beta H_{fSHO}}$. We can perform a quantum quench by making $\mu$ time dependent. For simplicity, let us consider that we take $\mu$ to zero suddenly.\footnote{But note that the following results apply to any functional form of $\mu$ as a function of time $t$. It can be slow quench, a zig-zag quench,...etc, or the final value of $\mu$ can be any real number.} So, $\mu(t)$ is a step function. The result is that as soon as $\mu$ goes to zero, the system is described by the thermal ensemble $e^{-\beta\eta {\bf Q}}$, where $\eta=\mu$. So, in summary
\begin{equation}
H_{fSHO},\, e^{-\beta H_{fSHO}} \quad \xrightarrow[\text{profile}]{\text{any quench}}\quad H=0,\, e^{-\beta\eta {\bf Q}},\, \eta=\mu \quad \text{instantaneously}\
\label{inst_therm_fSHO}
\end{equation}
The Green's functions change from oscillatory behaviour ($\sim ie^{-it\mu}$) to contant values ($\sim i$). It is a special kind of instantaneous thermalization where temperature remains constant but the chemical potential changes.\footnote{Note that this happens only when the initial state is a thermal state. Pure excited states would not exhibit this behaviour of instantaneous thermalization in the usual sense. However, one can consider generalized Gibb's ensembles with infinite number of conserved charges to cover pure excited states. But as mentioned above we will not be considering non-equilibrium dynamics of pure excited states in this work.} This still happens even when the initial Hamiltonian has other terms as long as ${\bf Q}$ is a conserved charge, meaning ${\bf Q}$ commutes with the other terms. In our case here, the other terms will be the SYK interaction terms. So far this does not explain the instantaneous thermalization in the dilute gas phase when we perform quantum quench using time dependent SYK term. The connection is that the dilute gas phase consists of only long-lived quasi-particle excitations at very low density. The quantum quench slightly changes the fuzzy dressing of the quasiparticles. It also leads to the change in the temperature during the quench process unlike the case of fermionic SHO above. This argument is supported by the fact that the spectral function changes only slightly during our quench process. On the other hand, the quench process in the liquid phase changes the spectral function significantly. Detailed comparison can be found in section \ref{subsec:instherm_int}.

The smallest time scale in our theory is $1/J_4=1$. We used discrete time step-size $dt$ smaller than this value. So, the instantaneous thermalization is not due to use of inadequate time scale in our numerics. Moreover, it has been found that the dilute gas phase of two-sites coupled SYK model does not thermalize instantaneously \cite{Sorokhaibam_2021}. We believe that this is due to the fact that the site hopping term does not commute with the SYK terms so the change in the dressing of the hopping term leads to a thermalization process which is not instantaneous. It is unlikely that dilute gas phase in higher dimensional theories would thermalize instantaneously. Mass quenches in higher dimensional free relativistic theories have non-trivial time dependence \cite{Mandal:2015kxi,Paranjape:2016iqs,Banerjee:2019ilw}.

We will also study the non-equilibrium dynamics of $(q=2,4)$ SYK model. The $q=2$ SYK term alone is integrable. The $j_{2,ij}$ couplings can be diagonalized resulting in a theory of free N fermions with random masses. There is no sharp phase transition if we consider the $(q=2,4)$ SYK model in the absence of chemical potential \cite{Sorokhaibam_2020}. The Lyapunov exponent is suppressed by the integrable interaction. We calculate the Lyapunov exponent of this system to high precision. But it does not sharply go to zero even when the $(q=2)$ interaction strength is very strong and at very low temperatures. We will also show that this system thermalizes exponentially fast. These observations imply that this system is always in the highly chaotic liquid phase without quasi-particle excitation.

For a chaotic system the spectral statistics is described by Wigner-Dyson (WD) statistics. For a generic integrable system, the spectral statistics is described by Poisson distribution. It has been shown that $(q=2)$ SYK interaction forces the spectral statistics towards Poisson distribution \cite{Garcia-Garcia:2017bkg}.\footnote{Note that there is no phase transition which is otherwise claimed in this paper. Other deformation of SYK model has also been shown to have no phase transition \cite{Nosaka:2019tcx}.} This exercise was done considering finite size systems and numerically diagonalizing the Hamiltonian. But as we have mentioned above, the system is always chaotic and always thermalizes exponentially fast. So, this is a shortcoming of the BGS conjecture. It is hard to quantify how much chaotic a system is or if the system is completely integrable. There has been other works in similar line where the spectral statistics is not WD statistics but the system is nevertheless highly chaotic \cite{Akutagawa:2020qbj}.

Technically, we show the exponentially fast thermalization process in the liquid phase by calculating the effective temperature during the non-equilibrium dynamics. We find that, without chemical potential, the effective temperature has a single bump before settling down to the final value. With non-zero chemical potential, besides the bump the effective temperature oscillates noticeably before settling down to the final value. The oscillation frequency depends on the chemical potential and the frequency cutoff used to calculate the temperature.
In case of the dilute gas phase, it is not possible to calculate the temperature. Even out of equilibrium, it is not possible to calculate the effective temperature. So, we have to employ a different technique to show the instantaneous thermalization. The details can be found in section \ref{subsec:instherm_int}.

The conserved charge is
\begin{equation}
\langle Q \rangle=-i\,G^<(0)\
\label{Q_def}
\end{equation}
It does not change during our quench processes using time-dependent SYK interaction terms. This is true in both the phases. This is because the time-dependent SYK terms commute with the charge operator. This implies that spectral asymmetry frequency (SAF), in the liquid phase, remains unchanged when we perform quantum quenches using time-dependent SYK interaction terms. SAF is the shift in the peak of the product of the retarded Green's function and advanced Green's function \cite{Sachdev:2015efa}
\begin{equation}
G^R(\omega)G^A(\omega)=\Phi(\omega-\omega_s)
\label{saf}
\end{equation}
$\Phi(\omega)$ is an even function of $\omega$. $\omega_s$ is also the position of the peak of the spectral function $A(\omega)=-2\,\text{Im} G^R(\omega)$. The relation between SAF and the conserved charge at low temperature is \cite{Sachdev:2015efa}
\begin{equation}
Q=\frac{1}{4}\left[3-\tanh(\beta\omega_s/q)\right]-\frac{1}{\pi}\tan^{-1}\left(e^{\beta\omega_s/q}\right)\
\label{Q_SAF}
\end{equation}
It is interesting that the conserved charge $Q$ depends only on SAF even at high temperature. Other quantities like chemical potential and temperature changed when we compare the final state and the initial state after such quantum quenches.

\vspace{0.5cm}

In summary, the technical results of this work are:
\begin{enumerate}
\item We show that both the highly chaotic liquid phase and the dilute gas phase of charged SYK model thermalize. A system in the dilute gas phase thermalizes instantaneously.
\item $(q=2,4)$ SYK model is always in the highly chaotic liquid phase without any quasi-particle excitation and the system always thermalizes exponentially fast. We also calculate the Lyapunov exponent of this system with high precision.
\item In quantum quenches starting from the liquid state, the effective temperature is non-monotonic.
\end{enumerate}

Our initial aim of this work was to show that a system in the dilute gas state thermalizes very slowly (or fails to thermalize completely) but instead we find that the system thermalizes instantaneously. Failure of thermalization in highly interacting systems is a topic of intense research interest in experimental as well as theoretical physics. There are two popular paradigms in which a chaotic system fails to thermalize. The first one is the existence of quantum scars \cite{PhysRevLett.53.1515,Bernien_2017,Turner2018WeakEB,PhysRevB.98.155134,Ho_2019,Khemani_2019,Lin_2019}. Quantum scars are eigenstates which violates eigenstate thermalization hypothesis (ETH). They also have finite energy  density but anomalously low entanglement \cite{Mukherjee_2020}. So states (pure or mixed) formed out of quantum scars do not thermalize. The other paradigm is many body localization (MBL) \cite{Pal_2010,Nandkishore_2015,Alet_2018,Abanin_2019}. MBL is the suppression of chaos and slowing down of thermalization of an otherwise chaotic system due to the introduction of random disorder.

It would be interesting to examine closely the nature of the dilute gas phase. We believe that an analytical treatment might be possible especially for this phase. The dynamics of weakly interacting classical systems are well understood. The most famous example being the Fermi–Pasta–Ulam problem where the system fails to thermalize for exponentially long times even when there is a small but finite non-linear term. Similar phenomenon in weakly interacting quantum systems called prethermalization has been a topic of great interest in recent times \cite{Marcuzzi_2013,Essler:2013gta,Howell_2019}.

Instantaneous thermalization has previously been shown in quantum quenches starting from the ground state of a gapped theory to a gapless theory in two spacetime dimensions \cite{Mandal:2015kxi,Paranjape:2016iqs}. This applied only to correlators consisting of holomorphic operators of the 2D conformal field theory. Other correlation functions, say, of scalar operators, do not thermalize instantaneously in general. In holographic set-up, it has also been shown that one-point functions thermalize instantaneously \cite{Bhattacharyya:2009uu} but two-point functions does not \cite{Balasubramanian:2011ur}. It is worth mentioning here that \cite{Ebrahim:2010ra} claims that Hamadard functions thermalize instantaneously. But again, this has been explained in \cite{Balasubramanian:2011ur} that this particular two-point function does not probe inside the apparent horizon during black hole formation, which is why they thermalize instantaneously. Other two-point functions on the boundary theory do not thermalize instantaneously. Recently, it has also been shown that $(q\to\infty)$ SYK model thermalizes instantaneously \cite{Eberlein:2017wah}.

We would like to point out the differences of our work from \cite{Haldar:2019slc} which considers a modified system consisting of SYK model coupled to quadratic peripheral fermions. This modified system has the familiar highly chaotic non-Fermi liquid (NFL) phase and a Fermi liquid (FL) phase. Interestingly the above paper considers the modified model without any mass or chemical potential. In this setting, the system is either in the NFL phase or in the FL phase depending on a parameter $p$ which is the ratio of the number of the quadratic peripheral fermions and the number of SYK fermions. So depending on the value of $p$, the Hamiltonian of the modified system dictates if the system is in the NFL phase or in the FL phase. This is different from our present case where for the same Hamiltonian (at a given temperature and a given chemical potential) the system can be either in the highly chaotic liquid phase or the dilute gas phase. Another important difference is that in the above paper the modified system in the FL phase thermalizes slowly but in our present work we observe that a system in the dilute gas phase thermalizes instantaneously.

\vspace{0.5cm}

The outline of this paper is as follows:
We will work out the details of SYK model with complex fermions in section \ref{sec:syk}. The Kadanoff-Baym (KB) equations are derived in \ref{subsec:kbeqns}. We also explain the numerical methods involved in solving the equations. In section \ref{sec:cSYK}, we explain the phase transition and the details of the two phases. We examine in details the non-equilibrium dynamics of the liquid state in section \ref{subsec:therm_chaotic}. We show that a system in the dilute gas phase thermalizes instantaneously in section \ref{subsec:instherm_int}. In section \ref{sec:therm_q24}, we examine the non-equilibrium dynamics of $(q=2,4)$ SYK model. Section \ref{conclusions} consists of conclusions from this work.

\section{SYK model with complex fermions}
\label{sec:syk}
The general Hamiltonian is given in (\ref{hsyk}). To make the derivation simpler we will first consider only $(q=4)$ SYK model with explicit time-dependence for the moment. The generalization to the full Hamiltonian of $(q=2,4,6)$ SYK model is straightforward. The action on the Schwinger-Keldysh contour $\mathcal{C}$ is
\begin{eqnarray}
S&=&\int_{\mathcal{C}} dt \left[\sum_{i=1}^{N}\psi_i^{\dagger}\left(i\partial_t-\mu\right)\psi_i-\sum_{i,j,k,l=1}^{N}j_{4,ij;kl}(t)\psi^{\dagger}_i\psi^{\dagger}_j\psi_k\psi_l\right]\
\label{fullaction}
\end{eqnarray}
$j_{4,ij;kl}(t)=j_{4R,ij;kl}(t)+ij_{4I,ij;kl}(t)$ are complex numbers. $j_{4R,ij;kl}(t)$ and $j_{4I,ij;kl}(t)$ are drawn from Gaussian distributions of zero mean and variances $J_4(t)$. Moreover,
\begin{equation}
j_{4,ij;kl}=j^*_{4,kl,ij}, \qquad j_{4,ij;kl}=-j_{4,ji,kl}, \qquad j_{4,ij;kl}=-j_{4,ij,lk}\
\end{equation}
We work with quenched averaging of the coupling in the large N limit \cite{Eberlein:2017wah,Bhattacharya:2018fkq}. The contour ordered Green's function is defined as
\begin{equation}
G(t,t')=-\frac{i}{N}\sum_{i=1}^N \mathcal{T_C}\left(\psi_i(t)\psi^{\dagger}_i(t')\right)\
\end{equation}
The partition function becomes
\begin{eqnarray}
Z&=&\int \mathcal{D}\psi^{\dagger} \mathcal{D}\psi \int \mathcal{D}G\,\mathcal{D}\Sigma \; \exp\left[-\int_{\mathcal{C}} dt \sum_i\psi^{\dagger}_i(\partial_t+i\mu) \psi_i\right.\nonumber\\
&&\qquad -\frac{N}{4}\int_{\mathcal{C}} dt_1 dt_2 J_4(t_1)J_4(t_2) G(t_2,t_1)^2G(t_1,t_2)^2\nonumber\\
&&\qquad \left.-i\int_{\mathcal{C}} dt_1 dt_2 \Sigma(t_1,t_2)\left\{G(t_2,t_1)+\frac{i}{N}\sum_i \psi_i(t_2)\psi^{\dagger}_i(t_1)\right\}\right]\
\label{part2}
\end{eqnarray}
Integrating out the fermions, the action in terms of the bilocal fields is
\begin{eqnarray}
S[G,\Sigma]&=&-i N \, \text{tr} \log \left[\partial_{t_1}\delta_{\mathcal{C}}(t_1,t_2)+i\mu\,\delta_{\mathcal{C}}(t_1,t_2)+i\Sigma(t_1,t_2)\right]\nonumber\\
&& \qquad+ \frac{i N }{4}\int_{\mathcal{C}} dt_1 dt_2 J_{4}(t_1)J_{4}(t_2) G(t_2,t_1)^2G(t_1,t_2)^2\nonumber\\
&&\qquad\qquad -\, iN\int_{\mathcal{C}} dt_1 dt_2 \Sigma(t_1,t_2)G(t_2,t_1)\
\label{effaction}
\end{eqnarray}
The equations of motion of $G$ and $\Sigma$ are
\begin{gather}
\label{SD1}
\left(i\partial_{t_1}+\mu\right)\delta_{\mathcal{C}}(t_1,t_2)-G(t_1,t_2)^{-1}=\Sigma(t_1,t_2)\\
\Sigma(t_1,t_2)= J_4(t_1)J_4(t_2)G(t_2,t_1)G(t_1,t_2)^2\
\label{SD2}
\end{gather}
These are the Schwinger–Dyson (SD) equations in the Schwinger-Keldysh contour. Generalizing these equations for the general Hamiltonian (\ref{hsyk}) of ($q=2,4,6$) SYK model, the SD equations are
\begin{eqnarray}
\label{genSD1}
&&\left(i\partial_{t_1}+\mu\right)\delta_{\mathcal{C}}(t_1,t_2)-G(t_1,t_2)^{-1}=\Sigma(t_1,t_2)\\
\Sigma(t_1,t_2)&=&J_2(t_1)J_2(t_2)G(t_1,t_2)+J_4^2 G(t_2,t_1)G(t_1,t_2)^2\nonumber\\
&&\qquad\qquad+J_6(t_1)J_6(t_2)G(t_2,t_1)^2G(t_1,t_2)^3\
\label{genSD2}
\end{eqnarray}
We obtain different Green's functions when we go from the Schwinger-Keldysh contour to the real time axis
\begin{equation}
G^>(t_1,t_2)=G(t_1^-,t_2^+), \qquad G^<(t_1,t_2)=G(t_1^+,t_2^-)\
\end{equation}
where $t^{\pm}=t\pm i\epsilon$. Operator at $t^+$ comes before operator at $t^-$. With this, the SD equations are
\begin{gather}
\label{rtSD1}
\frac{1}{i\omega-\mu-\Sigma_R(\omega)}=G_R(\omega)\\
\Sigma^{>(<)}(t_1,t_2)=J_2(t_1)J_2(t_2)G^{>(<)}(t_1,t_2)+J_4^2 G^{<(>)}(t_2,t_1)G^{>(<)}(t_1,t_2)^2\nonumber\\
\qquad\qquad+J_6(t_1)J_6(t_2)G^{<(>)}(t_2,t_1)^2G^{>(<)}(t_1,t_2)^3\
\label{rtSD2}
\end{gather}
where $G_R$ is defined in (\ref{GR}) and $\Sigma_R$ is also similarly defined. The equilibrium solution are solved numerically. The connection between the above two equations are the fluctuation-dissipation relations which gives the expression of $G^>(\omega)$ and $G^<(\omega)$ in terms of the spectral function $A(\omega)$.
\begin{gather}
G^>(\omega)=-i\,\frac{A(\omega)}{1+e^{\beta(\omega+\eta)}},
\qquad G^<(\omega)=i\,\frac{A(\omega)}{1+e^{-\beta(\omega+\eta)}}\\
A(\omega)=-2\,\text{Im}\, G_R(\omega)\
\end{gather}

In the absence of $\mu$ and $\eta$, the SD equations are same as that of Majorana SYK models \cite{Eberlein:2017wah,Bhattacharya:2018fkq}. This is because in thermal equilibrium,
\begin{eqnarray}
G^>(t_1,t_2)=-G^<(t_2,t_1)
\label{commajequiv}
\end{eqnarray}
This relation holds true even out-of-equilibrium during time evolution starting from thermal equilibrium. So in the absence of mass or chemical potential, quantum quenches using time-dependent $J_2, J_4$ or $J_6$ in SYK model with complex fermions are same as quenches with the same time-dependent couplings in SYK model with Majorana fermions. Also note that in thermal equilibrium,
\begin{equation}
G^{>(<)}(t_1,t_2)=-G^{>(<)}(t_2,t_1)^*\
\label{Gglmr}
\end{equation}
Again this relation also holds true out-of-equilibrium starting from thermal equilibrium. This relation halves the computer time for time evolution because we can solve $G^{>(<)}(t_1,t_2)$ for either only $t_1\geq t_2$ or $t_1 \leq t_2$.

We also verify conservation of energy during the quench processes. The expression for energy is
\begin{eqnarray}
\frac{E(t_1)}{N}&=&-i\mu G(t_1+i\epsilon,t_1)-i\int_{\mathcal{C}}dt_2\left[\frac{J_2(t_1)J_2(t_2)}{2}\,G(t_2,t_1)G(t_1,t_2)\right.\nonumber\\
&&\qquad\qquad +\frac{J_4^2}{4}\,G(t_2,t_1)^2G(t_1,t_2)^2+\left.\frac{J_6(t_1)J_6(t_2)}{6}\,G(t_2,t_1)^3G(t_1,t_2)^3\right]\nonumber\\
&=&-i\mu G^<(t_1,t_1)-i\int_{-\infty}^{t_1}dt_2\left[\frac{J_2(t_1)J_2(t_2)}{2}\,\left(G^<(t_2,t_1)G^>(t_1,t_2)-G^>(t_2,t_1)G^<(t_1,t_2)\right)\right.\nonumber\\
&&\qquad\qquad\qquad +\frac{J_4^2}{4}\,\left(G^<(t_2,t_1)^2G^>(t_1,t_2)^2-G^>(t_2,t_1)^2G^<(t_1,t_2)^2\right)\nonumber\\
&&\qquad\qquad\qquad \left.+\frac{J_6(t_1)J_6(t_2)}{6}\,\left(G^<(t_2,t_1)^3G^>(t_1,t_2)^3-G^>(t_2,t_1)^3G^<(t_1,t_2)^3\right)\right]\nonumber\\
\label{energy}
\end{eqnarray}

The temperature and chemical potential can be calculated from $G^>(t_1,t_2)$ and $G^<(t_1,t_2)$ using the relation
\begin{equation}
\frac{G^K(\omega)}{-2\,\text{Im} G^R(\omega)}=\tanh \left(\frac{\beta(\omega+\eta)}{2}\right)\
\label{tempchem}
\end{equation}
The effective temperature during the non-equilibrium time evolution is calculated by using the method in \cite{Eberlein:2017wah,Bhattacharya:2018fkq}. For this, we perform a coordinate transformation from $(t_1,t_2)$ to $t_+=t_1+t_2,t_-=t_1-t_2$. Fourier transform w.r.t. $t_-$ and using (\ref{tempchem}) at small $\omega$ region gives the effective temperature as a function of $t_+$. Note that $t_+$ increases or decreases in time step of $2\times dt$. Highly chaotic system without quasiparticles are expected to thermalize exponentially as a function of the final temperature where the thermalization rate is directly proportional to the final temperature. This indeed has been shown to be true for SYK model without chemical potential. The effective temperature is given by
\begin{equation}
T_{eff}(t_+)=T_f+\alpha e^{-\Gamma t_+},\qquad \Gamma=cT_f\
\label{thermrate}
\end{equation}
where $T_f$ is the final temperature.

\subsection{Kadanoff-Baym equations and quantum quenches}
\label{subsec:kbeqns}
We perform the quenches by solving the equations of motion of the Green's functions in integro-differential form which are well-known as Kadanoff-Baym (KB) equations. The details of the derivation of the KB equations from the SD equations can be found in \cite{Bhattacharya:2018fkq}. After performing a convolution in eqn (\ref{genSD1}) with $G$ we get
\begin{eqnarray}
i\partial_{t_1}G(t_1,t_2)=\mu\, G(t_1,t_2) +\int_{\mathcal{C}} dt_3 \Sigma(t_1,t_3)G(t_3,t_2)\\
-i \partial_{t_2} G(t_1,t_2)=\mu\, G(t_1,t_2)+\int_{\mathcal{C}} dt_3 G(t_1,t_3)\Sigma(t_3,t_2)\
\end{eqnarray}
The real time KB equations are obtained after contour deformation (Langreth rule). The imaginary leg in the contour is removed using Bogoliubov principle with weakening initial correlations.
\begin{eqnarray}
i\partial_{t_1}G^>(t_1,t_2)&=&\mu\,G^>(t_1,t_2)+\int_{-\infty}^{\infty} dt_3 \left[\Sigma^>(t_1,t_3)G^A(t_3,t_2)+\Sigma^R(t_1,t_3)G^>(t_3,t_2)\right]\nonumber\\
\label{KB11}
\,\\
-i \partial_{t_2} G^>(t_1,t_2)&=&\mu\,G^>(t_1,t_2)+\int_{-\infty}^{\infty} dt_3 \left[G^>(t_1,t_3)\Sigma^A(t_3,t_2)+G^R(t_1,t_3)\Sigma^>(t_3,t_2)\right]\nonumber\\
\label{KB12}
\,\\
i\partial_{t_1}G^<(t_1,t_2)&=&\mu\,G^<(t_1,t_2)+\int_{-\infty}^{\infty} dt_3 \left[\Sigma^R(t_1,t_3)G^<(t_3,t_2)+\Sigma^<(t_1,t_3)G^A(t_3,t_2)\right]\nonumber\\
\label{KB13}
\,\\
-i \partial_{t_2} G^<(t_1,t_2)&=&\mu\,G^<(t_1,t_2)+\int_{-\infty}^{\infty} dt_3 \left[G^R(t_1,t_3)\Sigma^<(t_3,t_2)+G^<(t_1,t_3)\Sigma^A(t_3,t_2)\right]\nonumber\\
\label{KB14}
\end{eqnarray}
Pure $(q=2)$ SYK model does not thermalize \cite{Bhattacharya:2018fkq}. We can see from the above equations that $(q=2)$ SYK model even in the presence of chemical potential or a mass term does not thermalize at all. The system freezes conpletely as soon as the time-dependent perturbation stopped. This is because the expressions on the right side of (\ref{KB11}) and (\ref{KB12}) (or (\ref{KB13}) and (\ref{KB14})) are same. So,
\begin{equation}
G^{>(<)}(t_1,t_2)=G^{>(<)}(t_1+dt,t_2+dt)\
\end{equation}
Moreover, the final configuration is not thermal. So, $(q=2)$ SYK model equilibrates instantaneously but does not thermalize.

Unlike the $(q=2)$ SYK model, quantum quenches in the presence of $(q=4)$ interaction are non-trivial. Using relation (\ref{Gglmr}), we will only use (\ref{KB12}) to evolve $G^>(t_1,t_2)$ and (\ref{KB14}) for $G^<(t_1,t_2)$. The most convenient forms of the equations for numerical applications are ($t_a>t_b$)
\begin{eqnarray}
\label{nKB1}
-i \partial_{t_a} G^>(t_b,t_a)&=&\mu\, G^>(t_b,t_a)+\int_{t_b}^{t_a} dt_3 G^>(t_b,t_3)\left[\Sigma^<(t_3,t_a)-\Sigma^>(t_3,t_a)\right]\nonumber\\
&&\quad +\int_{-\infty}^{t_b} dt_3 \left[G^>(t_b,t_3)\Sigma^<(t_3,t_a)-G^<(t_b,t_3)\Sigma^>(t_3,t_a)\right]\\
-i \partial_{t_a} G^<(t_b,t_a)&=&-\mu\, G^<(t_b,t_a)+\int_{t_b}^{t_a} dt_3 G^<(t_b,t_3)\left[\Sigma^<(t_3,t_a)-\Sigma^>(t_3,t_a)\right]\nonumber\\
&&\quad +\int_{-\infty}^{t_b} dt_3 \left[G^>(t_b,t_3)\Sigma^<(t_3,t_a)-G^<(t_b,t_3)\Sigma^>(t_3,t_a)\right]\
\label{nKB2}
\end{eqnarray}
Note that the second integrals are same in the two equations. These equations can be solved incrementally/causally using a Predictor-Corrector scheme. We use forward difference for the prediction and we use backward difference for the correction. In the presence of non-zero $\mu$, the forward difference and the backward difference must be strictly taken otherwise the numerical scheme fails to converge. The predicted value is
\begin{equation}
G^{>}(t_b,t_a+dt)=G^{>}(t_b,t_a)(1+i \mu \, dt) +i dt\, I_p(t_b,t_a)
\end{equation}
where $I_p(t_b,t_a)$ consists of the sum of the two integrals in (\ref{nKB1}). The correction is performed using the backward difference formula and simple mixing to achieve convergence.
\begin{equation}
G^{>}(t_b,t_a+dt)=\frac{G^{>}(t_b,t_a)(1+i\mu \,dt) +i dt\,I_p(t_b,t_a)}{2}+\frac{G^{>}(t_b,t_a) +i dt\,I_c(t_b,t_a+dt)}{2(1-i\mu\,dt)}\
\end{equation}
where $I_c(t_b,t_a+dt)$ is the sum of the integrals in (\ref{nKB1}) calculated using the predicted value of $G^{>}(t_b,t_a+dt)$. Similarly, $G^{<}(t_b,t_a+dt)$ is also calculated. For the diagonal term $G^{>(<)}(t_a,t_a)$, we use the sum of (\ref{KB11}) and (\ref{KB12}).
\begin{eqnarray}
G^{>}(t_a+dt,t_a+dt)&=&G^{>}(t_a,t_a)+idt\,I_{diag}(t_a)\\
G^{<}(t_a+dt,t_a+dt)&=&G^{<}(t_a,t_a)+idt\,I_{diag}(t_a)\\
I_{diag}(t_a)&=&\int_{-\infty}^{t_a} dt_3 \left[+G^>(t_a,t_3)\Sigma^<(t_3,t_a)-G^<(t_a,t_3)\Sigma^>(t_3,t_a)\right.\nonumber\\
&&\qquad \left.-\Sigma^>(t_a,t_3)G^<(t_3,t_a)+\Sigma^<(t_a,t_3)G^>(t_3,t_a)\right]\
\end{eqnarray}

An important check for our numerical codes is to perform equilibrium time evolution without quantum quenches. It serves three purposes. First it verifies that our codes are correct. It also verifies that our initial data are correct and accurate. The initial data are generated by solving the SD equations. Lastly, it verifies that numerical errors are under control.

\section{Charged SYK model}
\label{sec:cSYK}
\begin{figure}[h]
\begin{center}
\includegraphics[width=0.5\textwidth]{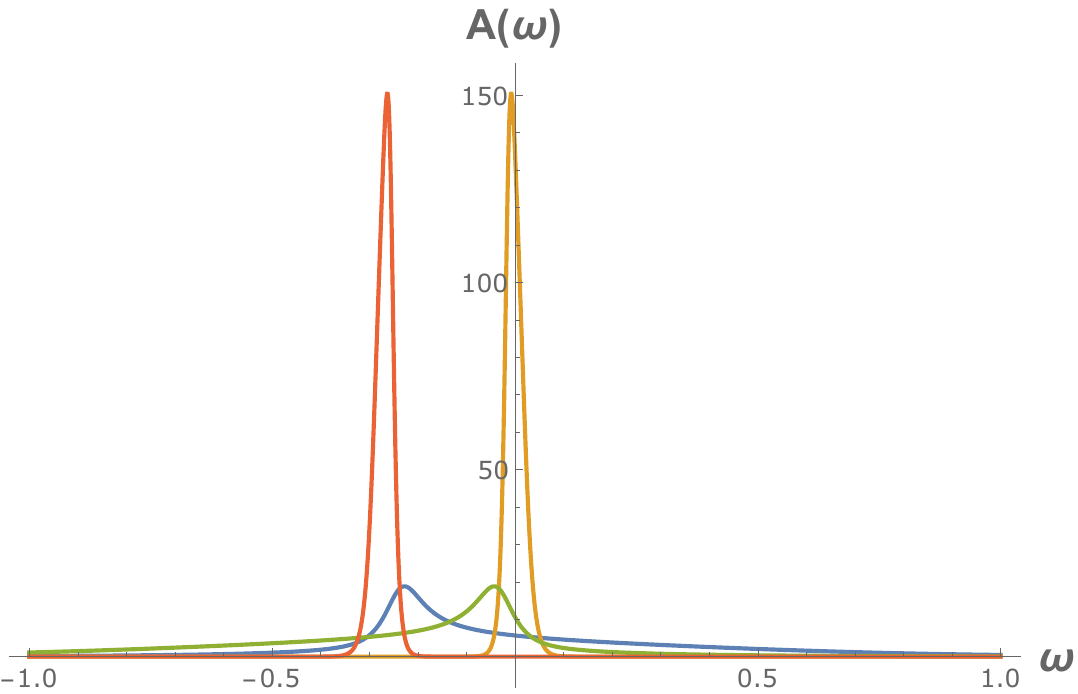}
\caption{\small Plots of spectral function $A(\omega)$ in the two phases with chemical potential and with mass term. The blue curve is for the liquid state with chemical potential $\eta=0.27$. The yellow curve is for the dilute gas state with $\eta=0.27$. The green curve is for the liquid phase with mass $\mu=-0.27$. The red curve is for the dilute gas phase with $\mu=-0.27$.}
\label{fig:spectral_func}
\end{center}
\end{figure}

In this section we will consider the system which has non-zero chemical potential or mass term in the Hamiltonian. We will consider only $(q=4)$ interaction. The Hamiltonian is \footnote{Note that later we will perform quantum quenches in this system. Keeping the $q=4$ SYK term unchanged, we will use time-dependent $q=2$ or $q=6$ SYK term to take the $(q=4)$ SYK system out of equilibrium. So in that context, the full time-dependent Hamiltonian will consist of $q=4$ SYK term and a time-dependent $q=2$ or $q=4$ SYK term.}
\begin{equation}
H=\mu \sum_i\Psi^{\dagger}_i\Psi_i+\sum_{i,j,k,l=1}^{N}j_{4,ij;kl}\Psi_i^{\dagger}\Psi_j^{\dagger}\Psi_k\Psi_l\
\end{equation}
As we have mentioned in section \ref{sec:intro}, the system can undergo a liquid-gas phase transition in the presence of effective chemical potential (explicit chemical potential $\eta$ or mass $\mu$ or both). In the highly chaotic liquid phase, the system does not have a quasiparticle picture. In the dilute gas phase, the system consists of long-lived quasi-particles at very low density. Figure \ref{fig:spectral_func} are plots of spectral functions in the two different phases. The spectral function in the dilute phase has a sharp single peak.

\begin{figure}[h]
\centering
\begin{subfigure}{.5\textwidth}
  \centering
  \includegraphics[width=.9\linewidth]{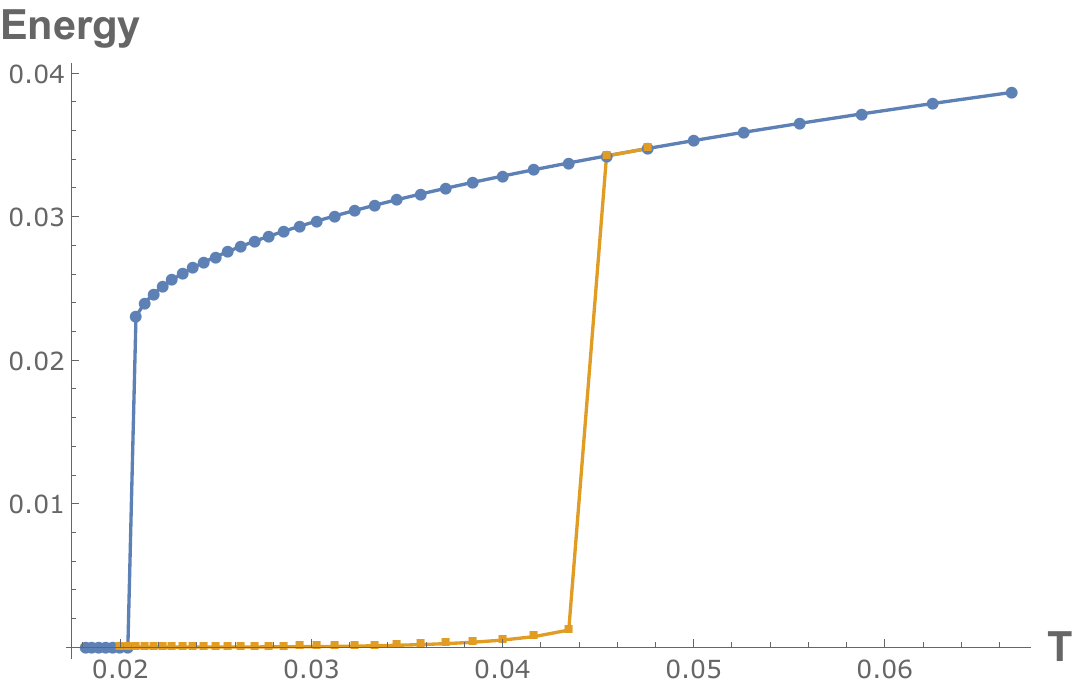}
  \caption{$\mu=0.27$, $\eta=0$}
  \label{fig:energy_mu}
\end{subfigure}%
\begin{subfigure}{.5\textwidth}
  \centering
  \includegraphics[width=.9\linewidth]{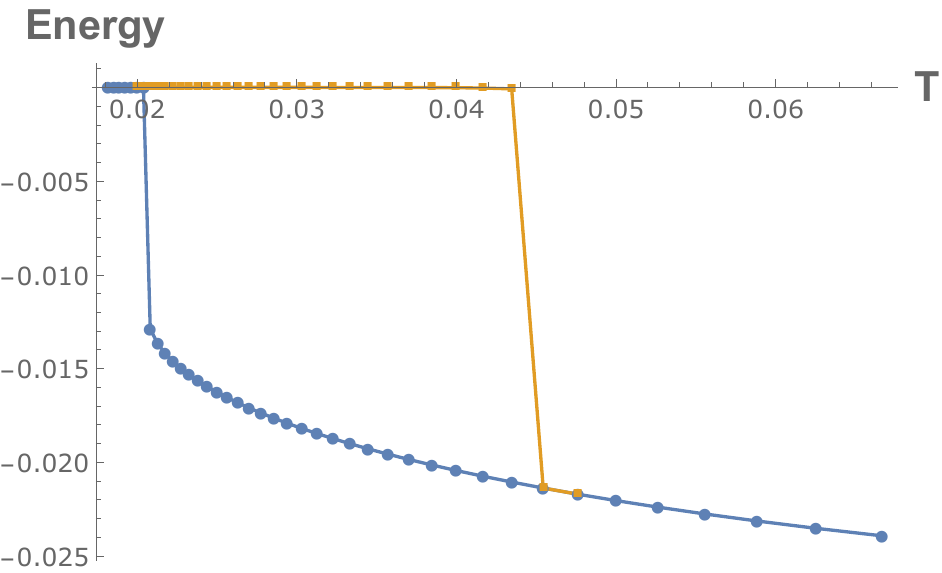}
  \caption{$\mu=0$, $\eta=0.27$}
  \label{fig:energy_eta}
\end{subfigure}
\caption{\small Plots of energy as a function of temperature T for the two different phases. The blue curves are for the liquid phase. The orange curves are for the dilute gas phase.}
\label{fig:energy}
\end{figure}
The two phases are separated by a large energy gap. Figure \ref{fig:energy} are plots of energy in the different phases with the mass term or the chemical potential.

The dilute gas phase also has occupation number close to $0$ or $1$ depending on the sign of the effective chemical potential. It can be easily seen from the plot of the spectral function. The area under the plot is equal to $1$, this is fixed by fermionic commutation relation. The occupation number is given by
\begin{eqnarray}
\frac{1}{N}\sum_i\langle \Psi^{\dagger}_i\Psi_i \rangle=\int d\omega\, \frac{A(\omega)}{1+e^{\beta(\omega+\mu+\eta)}}\
\end{eqnarray}
In case of the liquid phase, the spectral function is spreaded and the fermionic distribution function more effectively suppresses the occupation number.

Figure \ref{fig:phaselyap} is the plot of normalized Lyapunov exponent for the two different phases with varying temperature and a fixed chemical potential. Lyapunov exponent in the liquid phase is large. The normalized Lyapunov exponent increases as we decrease temperature for a fixed effective chemical potential. At the transition point from the liquid phase to the dilute gas phase, the Lyapunov exponent decreases sharply. But note that Lyapunov exponent in the dilute gas phase is non-zero and it is large at higher temperature.

\subsection{Thermalization in the liquid phase}
\label{subsec:therm_chaotic}
In this subsection and the next subsection, we will consider quantum quenches in $(q=4)$ SYK model in the presence of chemical potential. We will study time evolution of the system after the system is taken out of equilibrium using time-dependent $q=2$ or $q=6$ SYK interaction terms. We will consider only bump quenches where the time-dependent term is turned on for a short time duration and completely turned off again. The time-dependent Hamiltonians are
\begin{eqnarray}
H(t)&=&\mu \sum_i\Psi^{\dagger}_i\Psi_i+\sum_{i,j=1}^{N}j_{2,ij}(t)\Psi_i^{\dagger}\Psi_j+\sum_{i,j,k,l=1}^{N}j_{4,ij;kl}\Psi_i^{\dagger}\Psi_j^{\dagger}\Psi_k\Psi_l\\
\text{or} \qquad H(t)&=&\mu \sum_i\Psi^{\dagger}_i\Psi_i+\sum_{i,j,k,l=1}^{N}j_{4,ij;kl}\Psi_i^{\dagger}\Psi_j^{\dagger}\Psi_k\Psi_l\nonumber\\
&&\qquad \qquad +\sum_{i,j,k,l,m,n=1}^{N}j_{6,ijk;lmn}(t)\Psi_i^{\dagger}\Psi_j^{\dagger}\Psi_k^{\dagger}\Psi_l\Psi_m\Psi_n\\
j_{2,ij}(t)&=&j_{2,ij}f(t),\qquad\qquad j_{6,ijk;lmn}(t)=j_{6,ijk;lmn}f(t)\\
&& \text{where} f(t)= \begin{cases}
		1, & \quad 0<t<a\\
		0, & \quad \text{everywhere else.}\\
		\end{cases}\
\end{eqnarray}
Starting from the liquid state, we find that the system evolves non-trivially even after the time-dependent perturbation has been turned off. The system thermalizes exponentially fast.

The initial equilibrium state is prepared by solving the SD equations (\ref{rtSD1}) and (\ref{rtSD2}) without the $q=2$ and $q=6$ SYK terms. To perform the quantum quench, the KB equations (\ref{nKB1}) and (\ref{nKB2}) are solved numerically. The time dependent terms are turned on from $t_1,t_2=1 \times dt$ to $t_1,t_2=s \times dt$ where $dt$ is the discrete time step-size.

\begin{figure}[h]
\begin{center}
\includegraphics[width=0.5\textwidth]{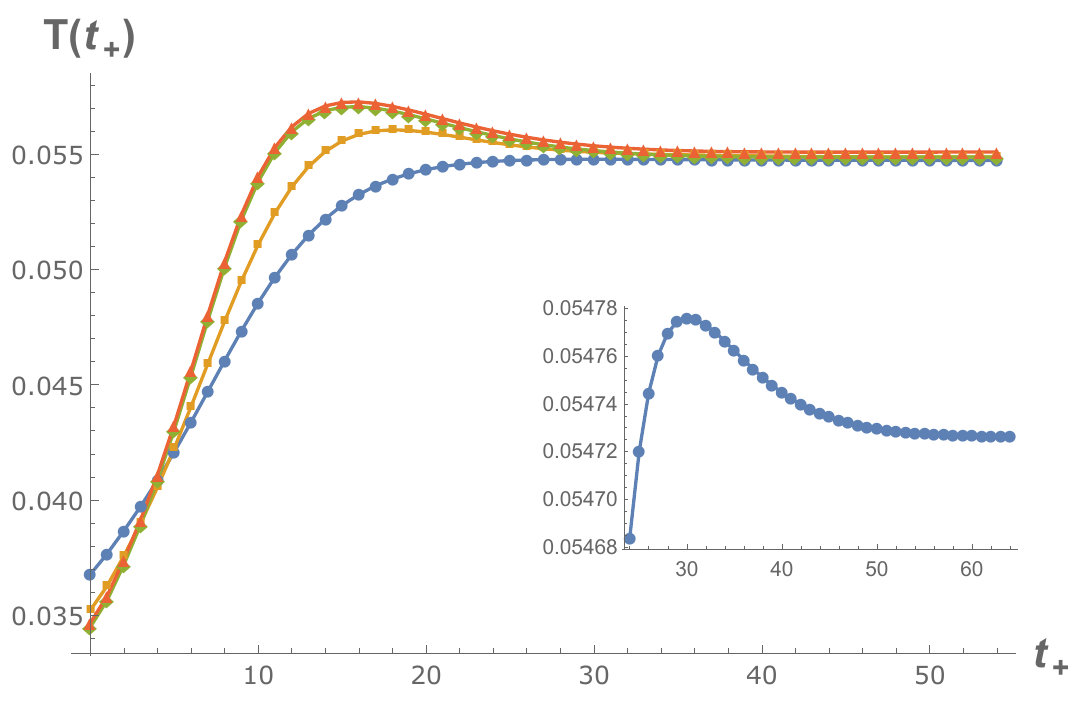}
\caption{\small {Plots of the effective temperature as a function of $\,t_+$ starting from different states and for different $\mu$. The quenches have been tuned so that the final inverse temperature is $\beta\sim 18$. All the quenches are performed with time dependent $q=4$ SYK interaction except for the red curve which was with time dependent $q=6$ SYK interaction. The blue curve is for $\mu=\eta=0$. The inset shows the non-monotonicity of the effective temperature for this case. The orange curve is for initial state with $\eta=0.24$. The green curve is for initial state $\eta=0$ but with $\mu=-0.263$. The red curve is for the same initial state as the green curve but the time dependent perturbation was $(q=6)$ interaction.}}
\label{fig:eff_temp}
\end{center}
\end{figure}

\begin{figure}[h]
\centering
\begin{subfigure}{.5\textwidth}
  \centering
  \includegraphics[width=.9\linewidth]{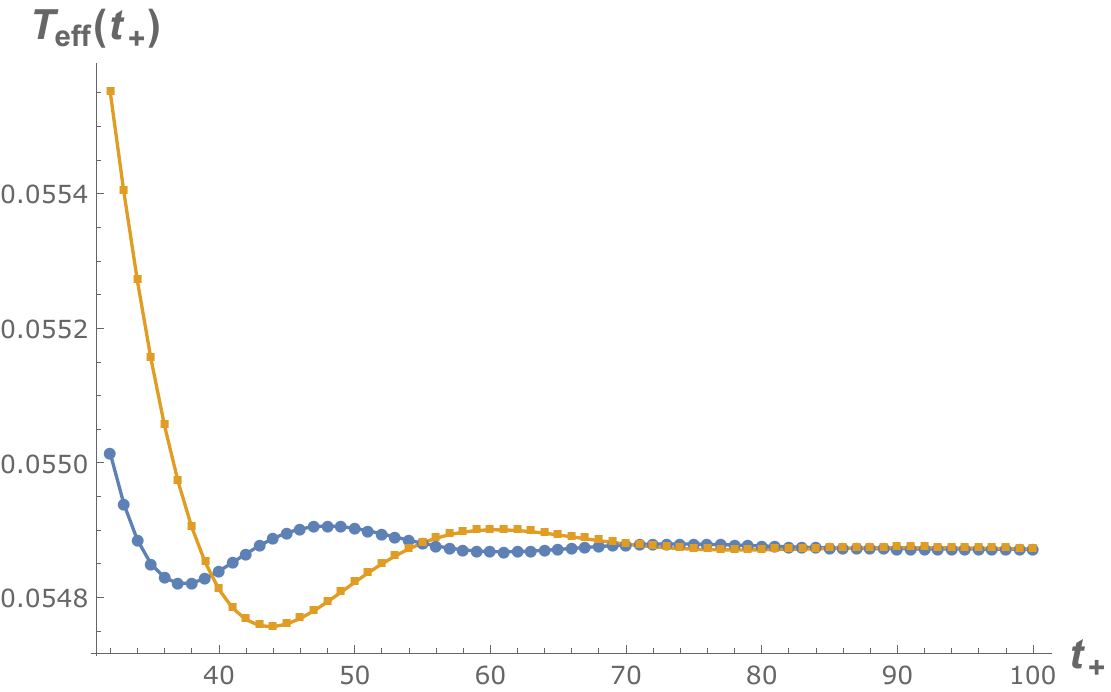}
  \caption{}
  \label{fig:temp_oscillation}
\end{subfigure}%
\begin{subfigure}{.5\textwidth}
  \centering
  \includegraphics[width=.9\linewidth]{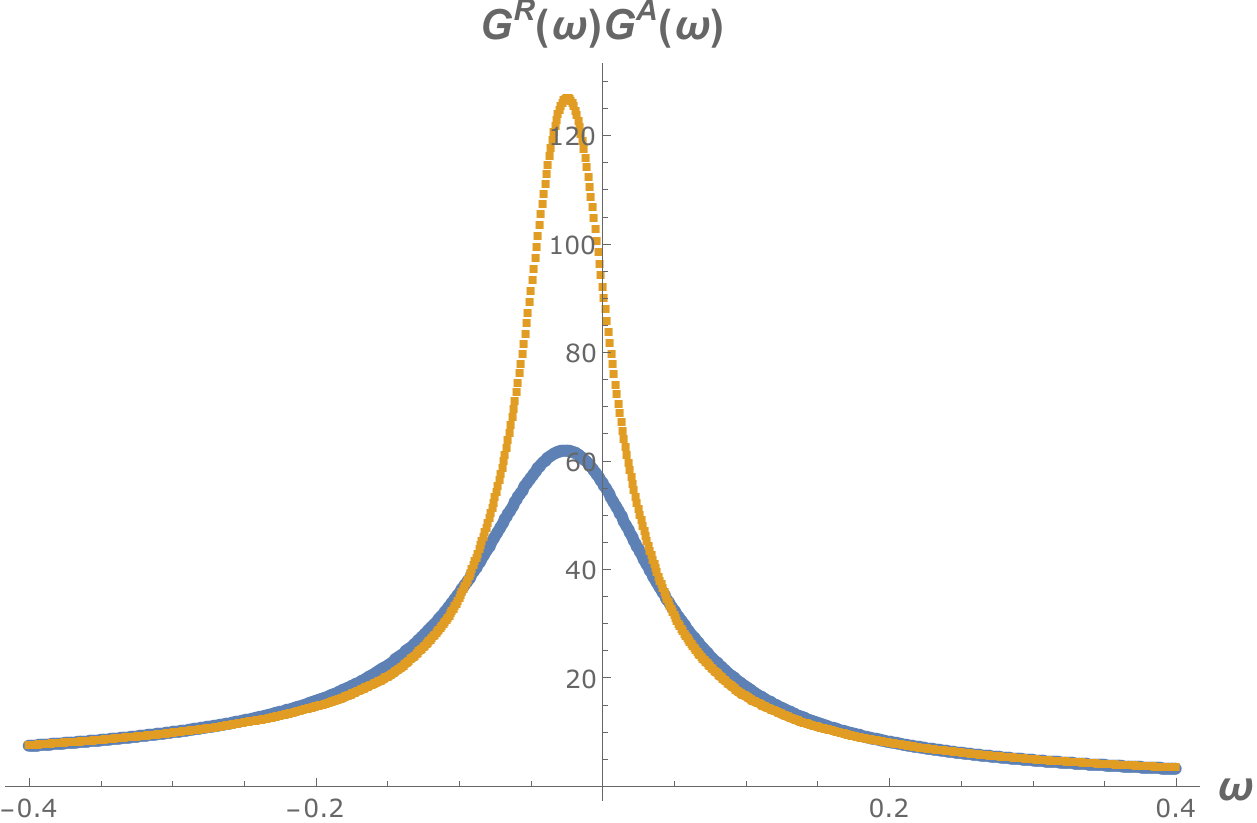}
  \caption{}
  \label{fig:saf}
\end{subfigure}
\caption{\small{(a) The effective temperature as a function of $\,t_+$ calculated using different frequency cutoffs. This is for the quench for $\mu=-0.263$ starting from $\beta_i=40, \eta=0$. The final state is $\beta_f=18.24, \eta=-0.029$. Note that the chemical potential changes during quench. (b) Considering the same quench, the spectral asymmetry frequency $\omega_s$ does not change during quantum quenches with SYK interactions. In this case, $\omega_s=-0.025$. The blue dots are $G^R(\omega)G^A(\omega)$ before quench and the orange dots are $G^R(\omega)G^A(\omega)$ after quench.}}
\label{fig:temp_os_saf}
\end{figure}

Figure \ref{fig:eff_temp} is the plot of effective temperatures for quenches in different settings but with the final temperature $T_f\sim 0.55$. As we can see the effective temperature is non-monotonic in all the cases. Without chemical potential, the effective temperature settles down to the final value. But with chemical potential, the effective temperature oscillates before settling down to the final value. As we have mentioned earlier, we use (\ref{tempchem}) at small $\omega$ region to calculate the temperature. The oscillation of the effective temperature actually depends on the frequency cutoff that we used for calculating the temperature. Figure \ref{fig:temp_oscillation} are the plots of effective temperature calculated using different frequency cutoffs.

Note that the value of the chemical potential changes during quenches. As mentioned in section \ref{sec:intro}, The charge operator commutes with the time-dependent SYK interaction terms. So, the conserved charge does not change during our quench processes. Accordingly, the spectral asymmetry frequency (SAF) also does not change during the quench processes. As defined in (\ref{saf}), SAF is the position of the maximum of $G^R(\omega)G^A(\omega)$. Figure \ref{fig:saf} are the plots showing the position of SAF before and after a quantum quench.

\begin{figure}[h]
\begin{center}
\includegraphics[width=0.5\textwidth]{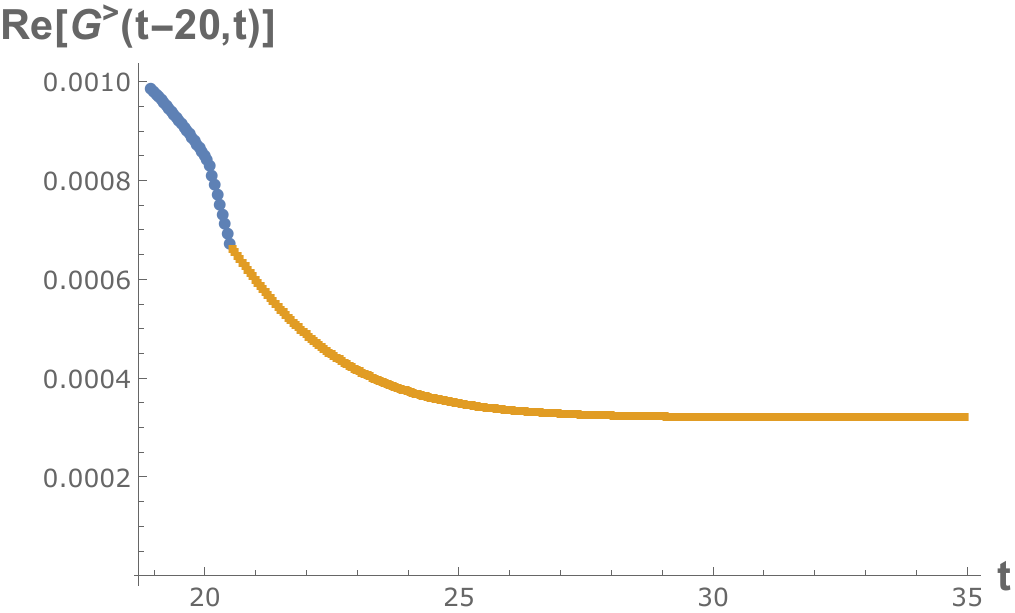}
\caption{\small{Plot of the real part of the greater Green's function $G^>(t-20,t)$ as a function of t for a quantum quench starting from the liquid state. The Green's function changes non-trivially even after both time argument have crossed the quench region (orange portion). The quench region is from $t=20.05$ to $t=20.50$.}}
\label{fig:ReGG_chaotic}
\end{center}
\end{figure}

In the liquid state, the Green's functions also thermalize exponentially fast. The Green's functions change non-trivially even after both the time arguments have crossed the quench region. The Green's functions converge towards their final values exponentially \cite{Bhattacharya:2018fkq}. Figure \ref{fig:ReGG_chaotic} is the plot of the real part of the greater Green's function $G^>(t-t_a,t)$ as a function of $t$. We will find that in case of quenches starting from the dilute gas state, the Green's functions will reach their final values abruptly.

\subsection{Thermalization in the dilute gas phase}
\label{subsec:instherm_int}
We will now consider quantum quenches where the initial state is a dilute gas state. With $\mu\neq 0$, the Green's functions oscillate since the effective theory is a weakly interacting massive theory. So, we have to consider very small time-step size $dt$ which results in very large number of discretization points. In case of $\mu= 0$ and $\eta\neq 0$, one can take large $dt$ but since the effective theory is a weakly interacting (almost) massless theory, the Green's function do not decay fast so one has to consider again a very large number of discretization points although $dt$ can be large.

Another technical difficulty while dealing with the dilute state is that the temperature cannot be calculated using (\ref{tempchem}) even for an equilibrium state solution calculated directly from the SD equations. The numerical precision is not good enough to cancel the spectral function and reproduce the $\tanh$ function. Consider the case of $\mu=0$ and $\eta=0.27$, as we can see in Figure \ref{fig:spectral_func} the spectral function is peaked around $\omega=0$. From (\ref{tempchem}), the expression on the left hand side is zero and rapidly varies only around $\omega=-\eta$. But around this region of frequency $\omega$, $G^K(\omega)$, $G^R(\omega)$ and $A(\omega)$ are numerically very small. So during the quench process, we will compare the energy of the final state with the energy of dilute gas states generated using the SD equations with different temperature and chemical potential. After finding a match in the energy, we compare $G^{>(<)}(t_1,t_2)$'s of the final state from the quantum quench and the thermal state.

As we have mentioned at the end of section \ref{subsec:kbeqns}, equilibrium time evolution without quantum quench is an important check. This check is very important for the dilute gas states due to the technical difficulties mentioned above.

We will present three cases:
\begin{enumerate}
\item For the dilute gas state with $\mu\neq 0$, we will consider $\mu=-0.27$, $\eta=0$ and $\beta = 30$. We used time step-size of $dt=0.03$ and 20001 discretization time steps from $\{-10000\times dt, 10000\times dt\}$. For the quantum quench, we perturb the system with $J_2=1$ from time $t=1\times dt$ to $t = 9 \times dt$. The system thermalizes to the dilute gas state with $\eta=-0.049172$ and $\beta=25.250$. The mass $\mu=-0.27$ is not changed during the quench.

\item For the dilute gas state with $\mu=0, \eta\neq 0$, we will consider $\mu=0$, $\eta=0.27$ and $\beta=50$ as the initial state. We used time step-size $dt=0.4$ and 20001 discretization time steps from $\{-10000\times dt, 10000\times dt\}$. For the quantum quench, we perturb the system with $J_2=0.1$ from time $t=1\times dt$ to $t = 9 \times dt$. The system thermalizes to the dilute gas state with $\eta=0.351023$ and $\beta=38.457$. The mass $\mu=0$ is not changed during the quench.

\item We also considered a case in which the system is perturbed with time-dependent $J_6$ term. For this, we used the same initial state as in the second case, i.e., $\mu=0$, $\eta=0.27$ and $\beta=50$. For the quantum quench, we perturb the system with $J_6=100000$ from time $t=1\times dt$ to $t = 9 \times dt$. The system thermalizes to the dilute gas state with $\eta=0.428934$ and $\beta=31.471$. The mass $\mu=0$ is not changed during the quench.
\end{enumerate}

\begin{figure}
\begin{subfigure}{.5\linewidth}
\includegraphics[width=.9\linewidth]{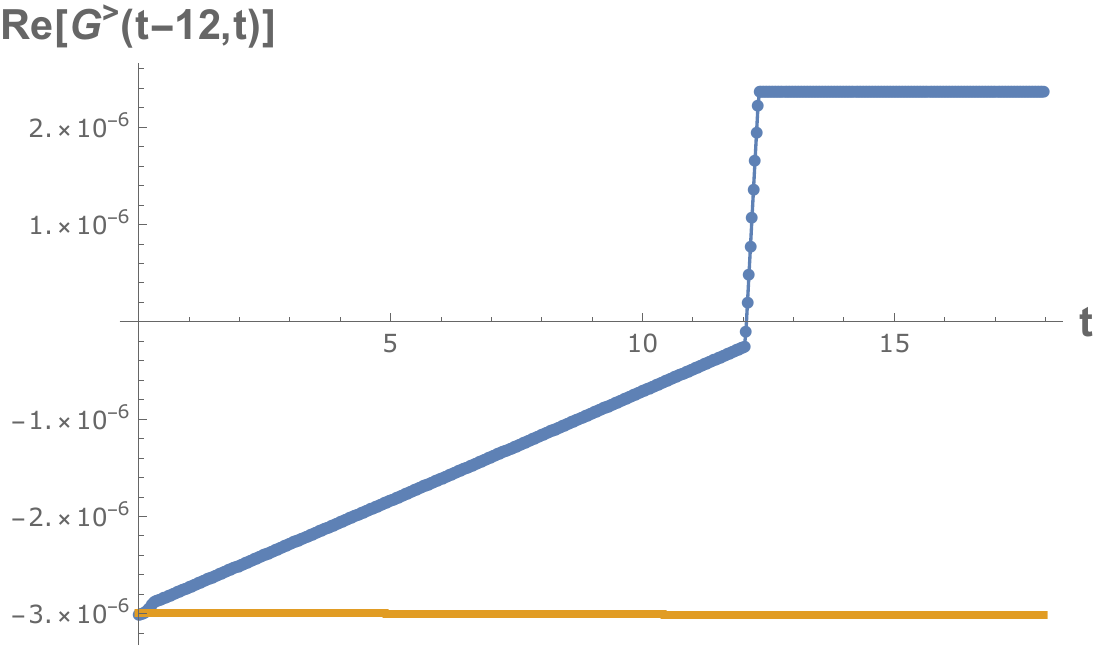}
\caption{}
\label{fig:sub1}
\end{subfigure}%
\begin{subfigure}{.5\linewidth}
\includegraphics[width=.9\linewidth]{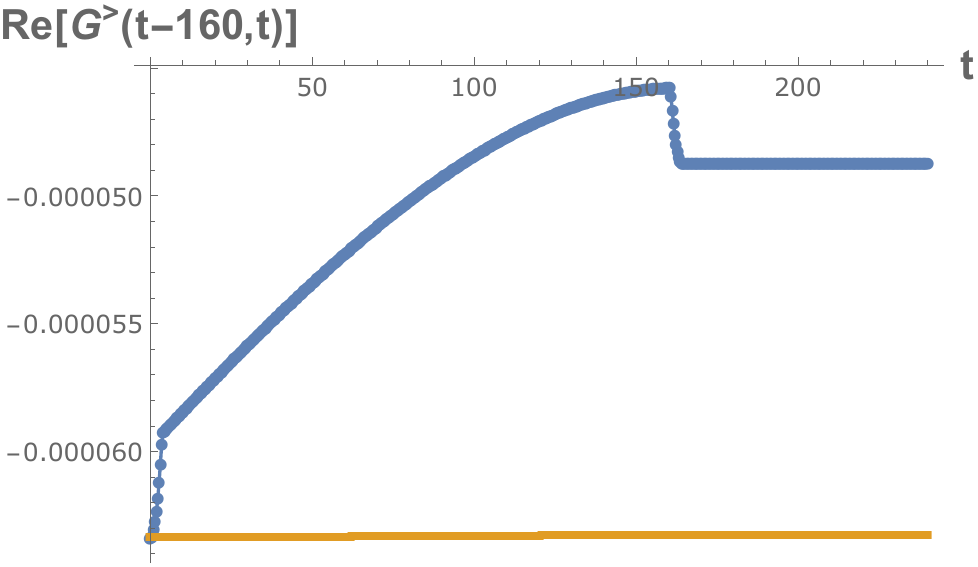}
\caption{}
\label{fig:sub2}
\end{subfigure}\\[1ex]
\begin{subfigure}{\linewidth}
\centering
\includegraphics[width=.5\linewidth]{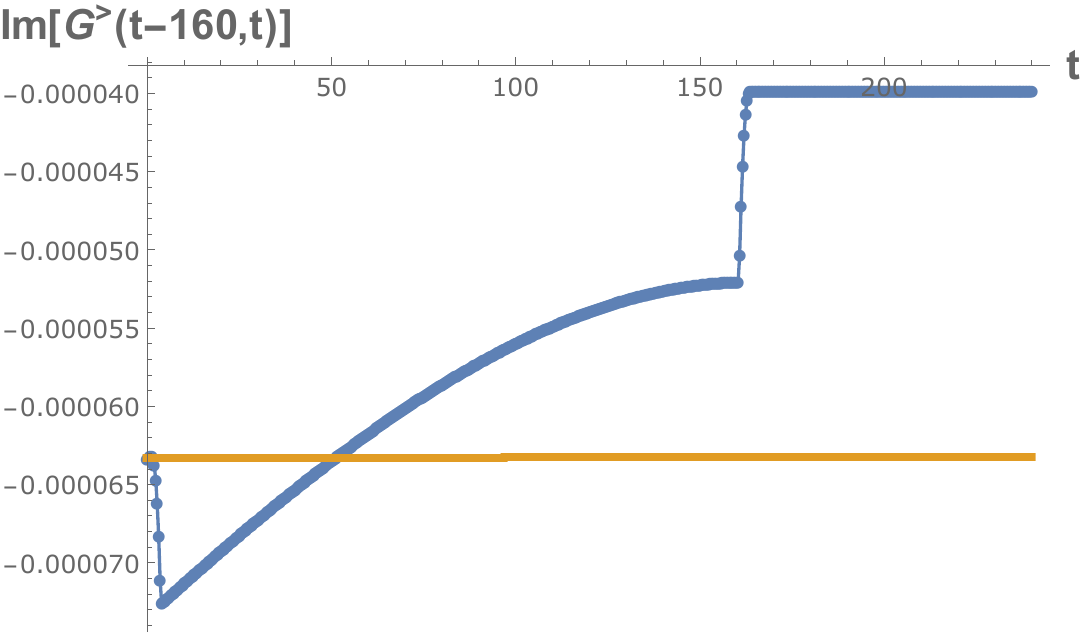}
\caption{}
\label{fig:sub3}
\end{subfigure}
\caption{\small{Time evolution of the real part of greater Green's function $G^>(t-t_a,t)$ as a function of $t$ for fixed $t_a$ showing the instantaneous thermalization (blue plots): (a) case 1 starting from the dilute gas state with $\mu=-0.27$, $\eta=0$ and $\beta=30$ and quantum quench using time dependent $J_2$ term, (b) case 2 starting from the dilute gas state with $\mu=0$, $\eta=0.27$ and $\beta=30$ and quantum quench using time dependent $J_2$ term, and (c) case 3 starting from the dilute gas state with $\mu=0$, $\eta=0.27$ and $\beta=30$ and quantum quench using time dependent $J_6$ term. The orange plots are equilibrium time evolutions for the respective initial states without any time dependent perturbations.}}
\label{fig:greens_int}
\end{figure}

We find that the systems stop evolving instantaneously when the time-dependent perturbation is turned off. The Green's functions freeze as soon as the two time arguments $t_1$ and $t_2$ cross the quench region. Figure \ref{fig:greens_int} are plots of the real parts of the greater Green's function $G^>(t-t_a,t)$ for different cases that we are considering. The imaginary part of $G^>(t-t_a,t)$ as well as the lesser Green's functions also freeze as soon as the two time arguments cross the quench region.

\begin{figure}
\centering
\begin{subfigure}{.5\textwidth}
  \centering
  \includegraphics[width=.9\linewidth]{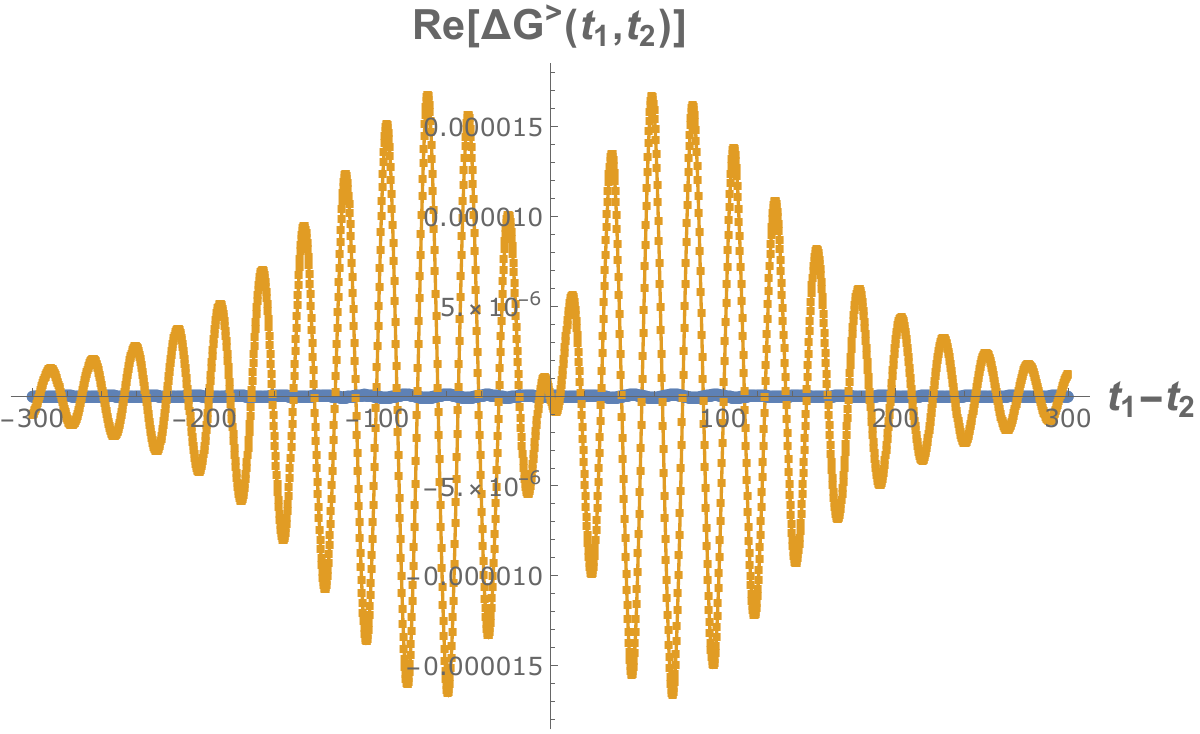}
  \caption{}
  \label{fig:therm_deltaReGGmu_Jt1}
\end{subfigure}%
\begin{subfigure}{.5\textwidth}
  \centering
  \includegraphics[width=.9\linewidth]{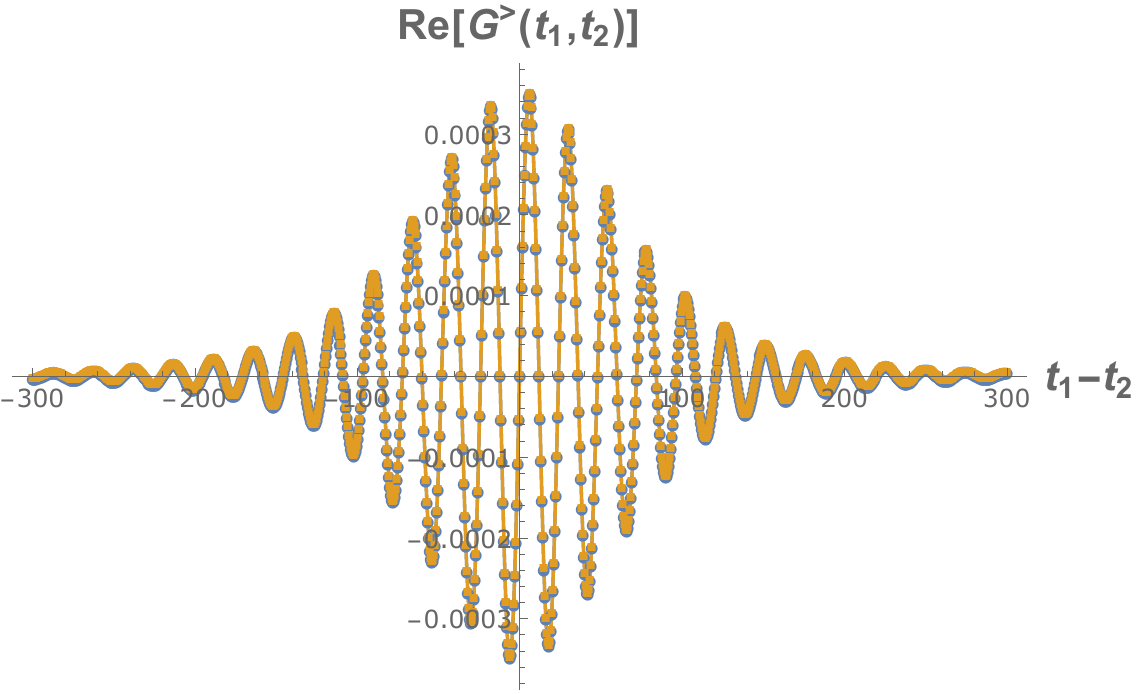}
  \caption{}
  \label{fig:therm_ReGGmu_Jt1}
\end{subfigure}
\begin{subfigure}{.5\textwidth}
  \centering
  \includegraphics[width=.9\linewidth]{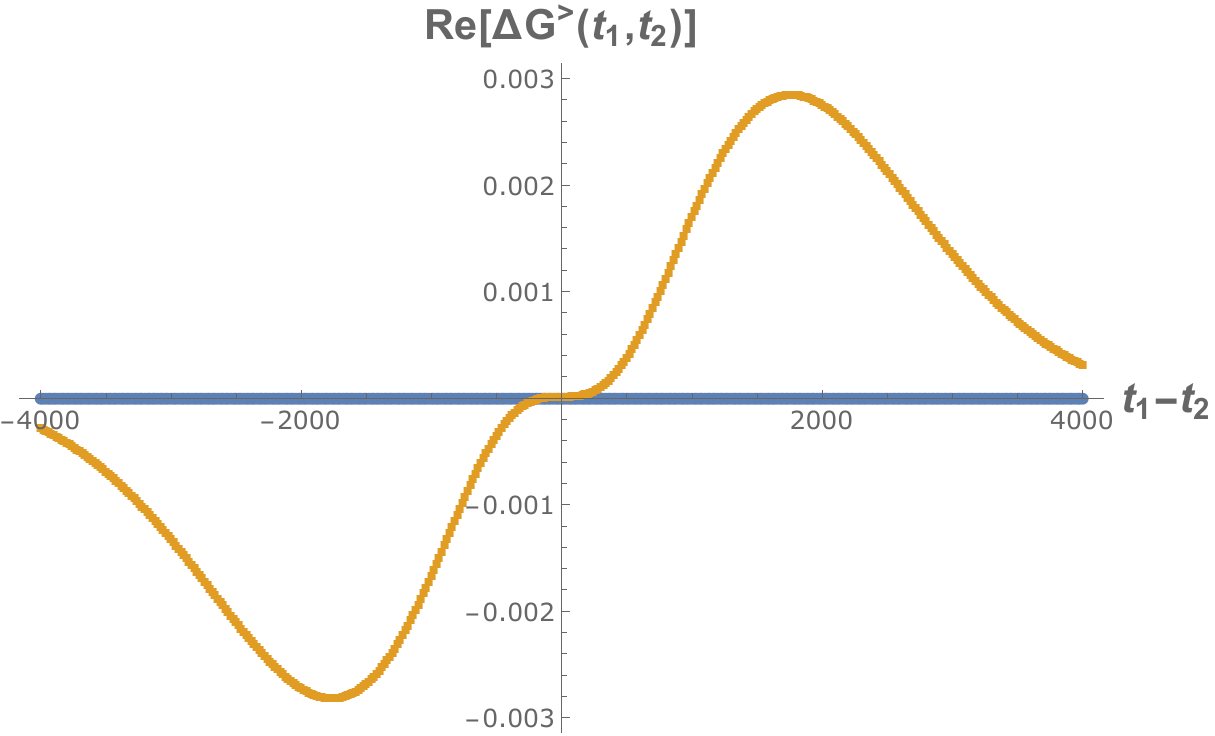}
  \caption{}
  \label{fig:therm_deltaReGGeta_Jt01}
\end{subfigure}%
\begin{subfigure}{.5\textwidth}
  \centering
  \includegraphics[width=.9\linewidth]{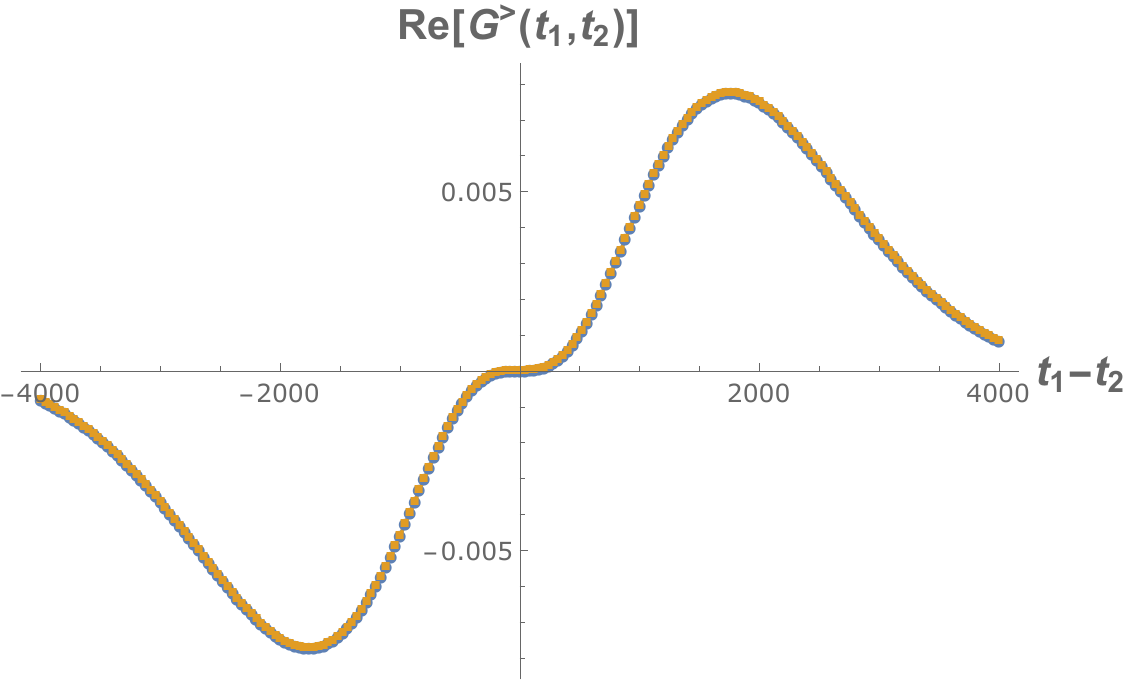}
  \caption{}
  \label{fig:therm_ReGGeta_Jt01}
\end{subfigure}
\begin{subfigure}{.5\textwidth}
  \centering
  \includegraphics[width=.9\linewidth]{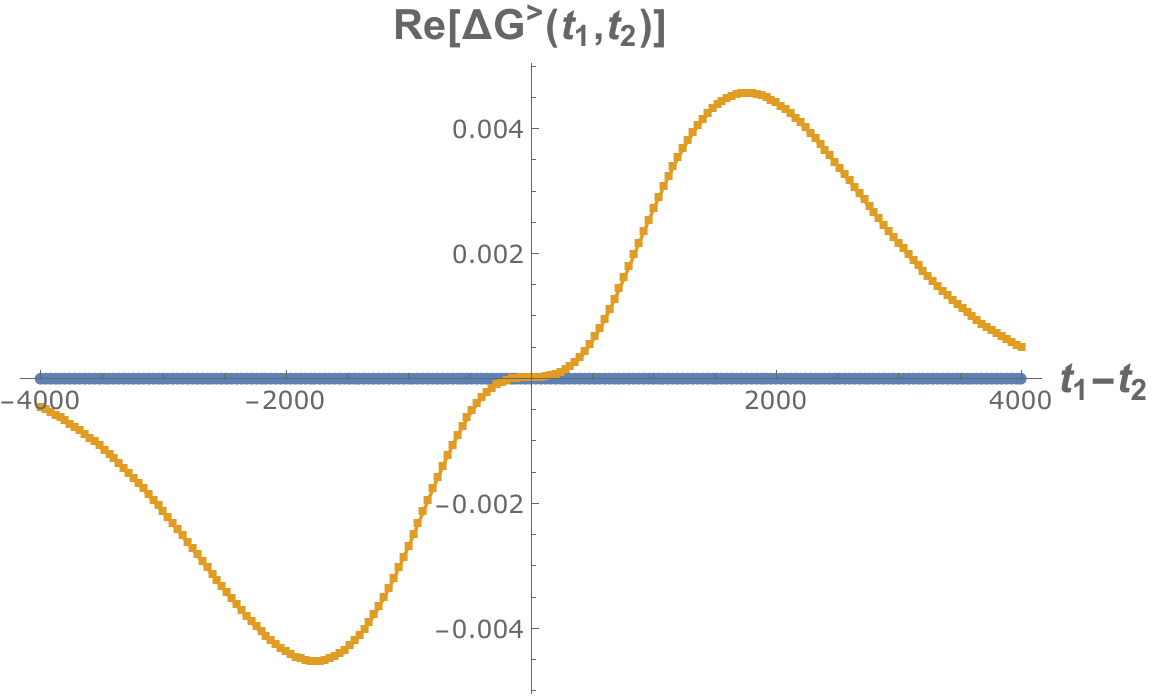}
  \caption{}
  \label{fig:therm_deltaReGGeta_Js100000}
\end{subfigure}%
\begin{subfigure}{.5\textwidth}
  \centering
  \includegraphics[width=.9\linewidth]{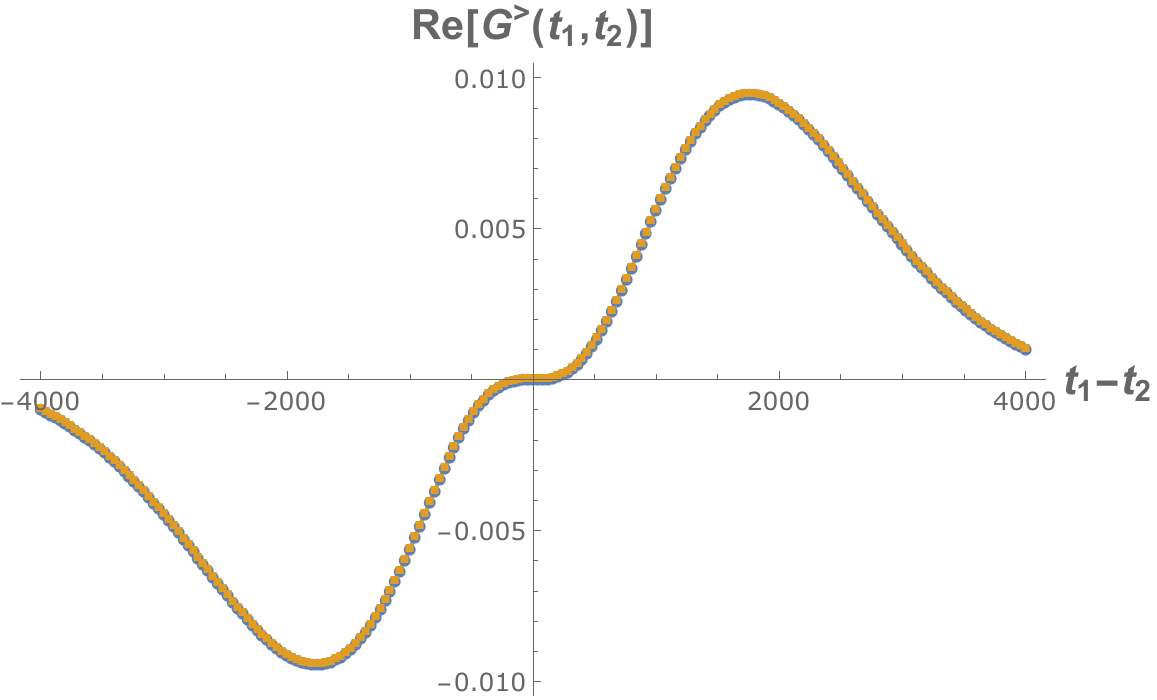}
  \caption{}
  \label{fig:therm_ReGGeta_Js100000}
\end{subfigure}
\caption{\small Plots on the left hand side: Blue curves are difference of the real part of the greater Green's functions of the final state and the thermal state with matching energy. Orange plots are difference of the real part of the greater Green's functions of the final state and the initial state. Plots on the right hand side: Blue curves are real part of the greater Green's functions of the final state. Orange plots are the real part of the greater Green's functions of the thermal state with matching energy. (a) and (b) are plots for case 1 starting from the dilute gas state with $\mu=-0.27$, $\eta=0$ and $\beta=30$ and quantum quench using time dependent $J_2$ term. (c) and (d) are plots for case 2  starting from the dilute gas state with $\mu=0$, $\eta=0.27$ and $\beta=30$ and quantum quench using time dependent $J_2$ term. And  (e) and (f) are plots for case 3 starting from the dilute gas state with $\mu=0$, $\eta=0.27$ and $\beta=30$ and quantum quench using time dependent $J_6$ term.}
\label{fig:therm_ReGG}
\end{figure}

We could not calculate the effective temperature and the effective chemical potential of the final state.\footnote{It is not clear a priori if the final state is thermal or not. What we meant is that we could not calculate the left hand side of (\ref{tempchem}).} So, we search for thermal states with energy equal to that of the final state from the quantum quench. This is done by solving the SD equations for different values of temperature and chemical potential. In this two dimensional parameter space, we could find a line for which the energy matches the energy of the final state. Then we compare the Green's functions of these thermal states with the Green's functions of the final state. Again we could find a unique thermal state for which the Green's functions match those of the final state. So, in conclusion, the system in the dilute gas state thermalizes instantaneously. The instantaneous thermalization happens irrespective of the perturbing term. We used time-dependent $J_2$ or $J_6$. Figure \ref{fig:therm_ReGG} are the plots of the real parts of the greater Green's functions $G^>(t-t_a,t)$ obtained after quantum quenches and thermal states generated using SD equations. It is evident that the final states are thermal states.

\begin{figure}
\centering
\begin{subfigure}{.5\textwidth}
  \centering
  \includegraphics[width=.9\linewidth]{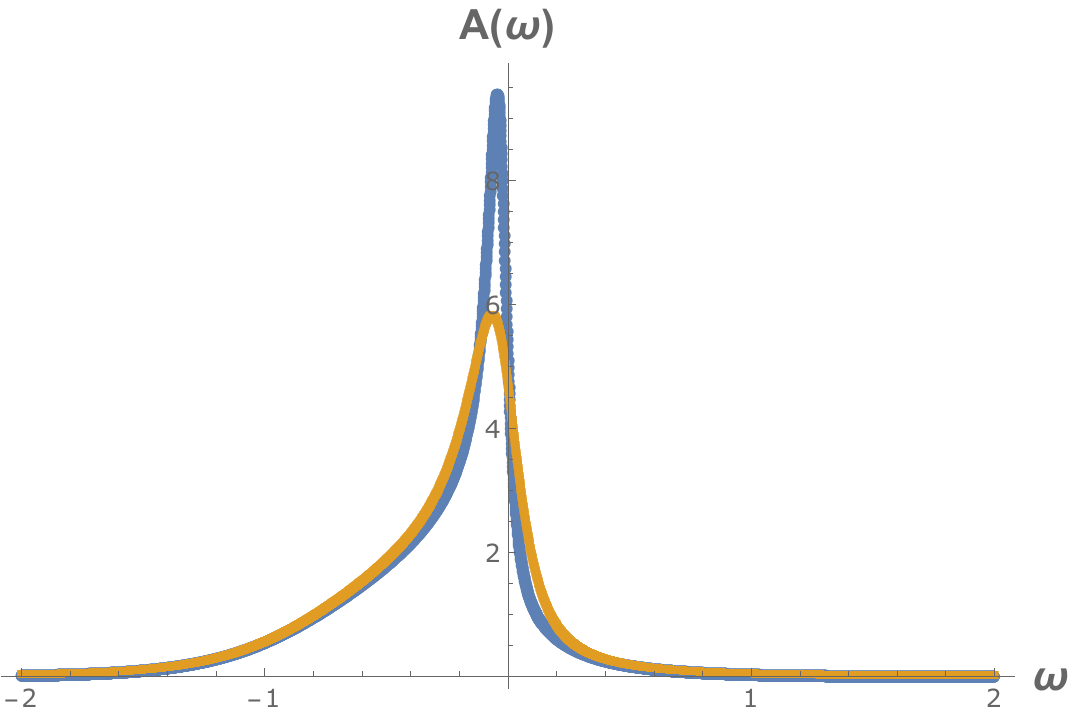}
  \caption{}
  \label{fig:delspec_chaotic}
\end{subfigure}%
\begin{subfigure}{.5\textwidth}
  \centering
  \includegraphics[width=.9\linewidth]{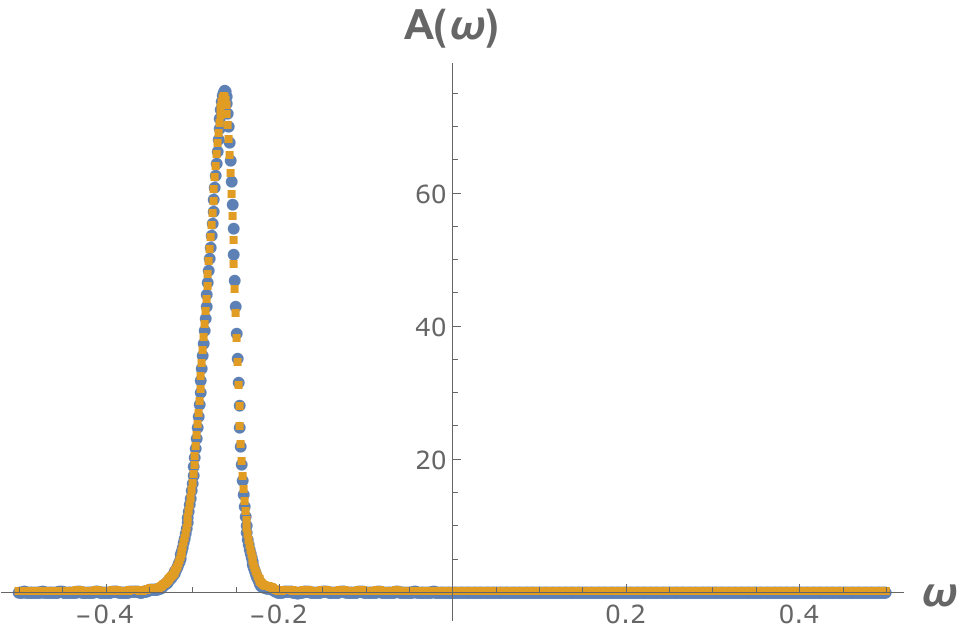}
  \caption{}
  \label{fig:delspec_dilute}
\end{subfigure}
\caption{\small{(a) The spectral function of the initial state(blue) and the final state(orange) in the liquid phase. (b)  The spectral function of the initial state(blue) and the final state(orange) in the dilute gas phase.}}
\label{fig:spec_func_chaotic_int_Inst}
\end{figure}

As we have pointed out in section \ref{sec:intro}, the spectral function changes only slightly during the quench process in the dilute gas phase. Figure \ref{fig:spec_func_chaotic_int_Inst} are plots of the spectral functions of the initial state and the final state in the two different phases. As, we can see the spectral function changes significantly during the quench process in the chaotic phase.

Now we will explain why the temperature of the dilute gas phase also changes during the quench process unlike in the case of fermionic SHO in (\ref{inst_therm_fSHO}). Since, the spectral function has a sharp peak near $\mu$ in the dilute gas phase, the fermionic distribution function is close to 1 at the energy scale where the non-trivial physics in happening. For example,
\begin{equation}
1/(1+e^{\beta \omega})\sim 0.9999, \quad \text{around} \quad \omega=-0.27, \beta\sim 30\
\end{equation}
This is the same reason why we could not calculate the temperature in the dilute gas phase using (\ref{tempchem}). Hence, the temperature in this phase is almost fixed by the spectral function itself. During the quench process when the spectral function changes (even though slightly) it leads to a change in the temperature of the system.

%
%
%

\section{$(q=2,4)$ SYK model}
\label{sec:therm_q24}
\begin{figure}[t]
\begin{center}
\includegraphics[width=0.6\textwidth]{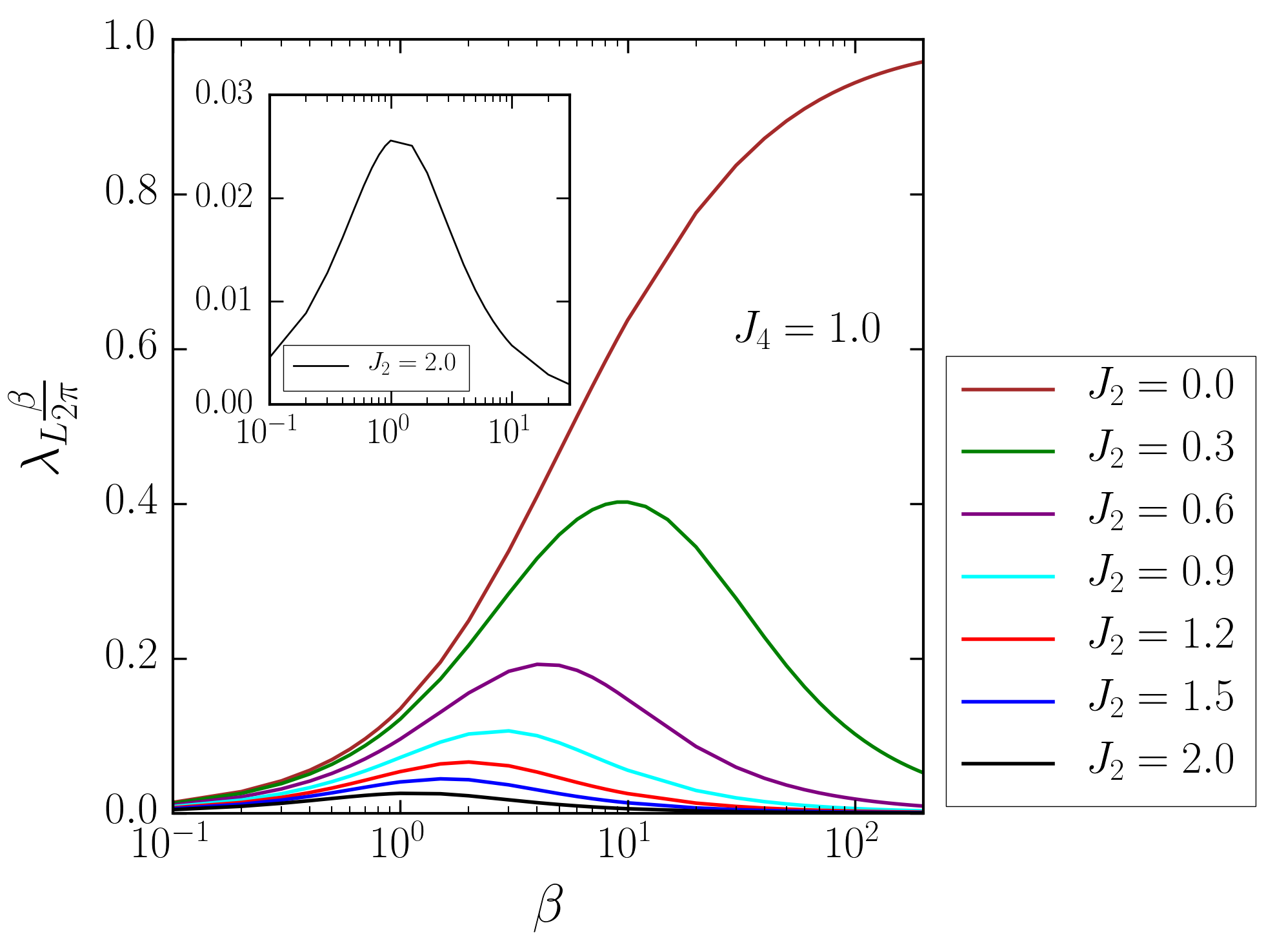}
\caption{\small{Plots of the normalized Lyapunov exponent as a function of inverse temperature $\beta$ for different values of $J_2$. For all the plots, $J_4=1$.}}
\label{fig:lyap_j2}
\end{center}
\end{figure}

In this section, we will consider $(q=2,4)$ SYK model without chemical potential. The Hamiltonian is \footnote{For non-equilibrium study, we will use time-dependent $q=6$ SYK term to take this system out of equilibrium.}
\begin{equation}
H=\sum_{i,j=1}^{N}j_{2,ij}\Psi_i^{\dagger}\Psi_j+\sum_{i,j,k,l=1}^{N}j_{4,ij;kl}\Psi_i^{\dagger}\Psi_j^{\dagger}\Psi_k\Psi_l\
\end{equation}
The $(q=2,4)$ SYK model is always in the liquid state. The system does not undergo liquid-gas transition. But the $(q=2)$ interaction suppresses chaotic nature of the $q=4$ SYK term. This can be seen from Figure \ref{fig:lyap_j2}. The Lyapunov exponent is greatly suppressed due to the presence of non-zero $J_2$ coupling but it does not sharply drop to a negligible value which is expected for a phase transition as in Figure \ref{fig:phaselyap}. The suppression of chaos has also been shown from spectral correlation calculation in \cite{Garcia-Garcia:2017bkg}. The introduction of large $J_2$ coupling forces the spectral statistics towards Poisson statistics. The spectral statistics obeys Poisson statistics for a generic integrable system while it obeys Wigner-Dyson(WD) statistics for a chaotic system. But this analysis are performed in finite systems (fixed N of the order of 10) so it is rather hard to precisely identify if the system is fully integrable or chaotic. In case of $(q=2,4)$ SYK model, the system is always in the highly chaotic liquid phase. In this work, we study the non-equilibrium dynamics of this system and show that the system thermalizes exponentially fast which is expected for a chaotic system. We also calculated the Lyapunov exponent of this system with high precision.

To calculate the Lyapunov exponent we followed the steps in \cite{Garcia-Garcia:2017bkg}. Note that in the absence of chemical potential, SYK model with $N$ complex fermions is same as SYK model with $2N$ Majorana fermions in every respect at large N limit. So, we performed the Lyapunov exponent calculations with Majorana fermions. $\mathcal{F}_1(t_1,t_2)$ from (\ref{four_pt}) satisfies
\begin{gather}
\mathcal{F}_1(t_1,t_2)=\int dt_3 dt_4 K_R(t_1,t_2,t_3,t_4)\mathcal{F}_1(t_3,t_4)\\
K_R(t_1,t_2,t_3,t_4)=G_R(t_1)G_R(t_2)\left[J^2_2+3J^2_4 G^2_{lr}(t_3-t_4)\right]\
\end{gather}
where $G_R(t)$ is the retarded Green's function and $G_{lr}(t)$ is $G^<(t+i\beta/2)$. Using the ansatz $\mathcal{F}_1(t_1,t_2)=e^{\lambda_L(t_1+t_2)/2}f(t_1-t_2)$ and going to the frequency domain using Fourier transforms, we get
\begin{equation}
f(\omega)=G_R(\omega+i\frac{\lambda_L}{2})G_R(\omega-i\frac{\lambda_L}{2})\left[J^2_2f(\omega)+3J^2_4\int\frac{d\omega'}{2\pi}\,G_{lr}(\omega-\omega')f(\omega')\right]\
\label{lyap_cal}
\end{equation}
Tuning $\lambda_L$ numerically to satisfy (\ref{lyap_cal}) gives the Lyapunov exponent. It is well-known that the Lyapunov exponent is bounded by $2\pi/\beta$. So, in the figures we have plotted normalized Lyapunov exponent instead which is defined as
\begin{equation}
\lambda^*_L=\lambda_L/(2\pi/\beta)
\end{equation}
Figure \ref{fig:lyap_j2} is the plot of the normalized Lyapunov exponent as a function of the inverse temperature for different values of $J_2$. It is interesting that with increasing inverse temperature the normalized Lyapunov exponent decreases gradually after a peak while in the presence of chemical potential the normalized Lyapunov exponent increases gradually in the liquid phase and sharply drops to a negligible value for the dilute gas phase as shown in Figure \ref{fig:phaselyap}.

To perform the quantum quenches, we consider two sets of the parameters which were considered in \cite{Sorokhaibam_2020}. We consider $J_2=0.5, J_4=1$ and show thermalization below $\beta=55$. Both the initial inverse temperature and the final inverse temperature are below $\beta=55$. We also consider $J_2=2, J_4=1$ and show thermalization below $\beta=15$. For these sets of parameters, it has been shown that the spectral statistics is almost completely Poisson statistics \cite{Garcia-Garcia:2017bkg}.

\begin{figure}[t]
\centering
\begin{subfigure}{.5\textwidth}
  \centering
  \includegraphics[width=.9\linewidth]{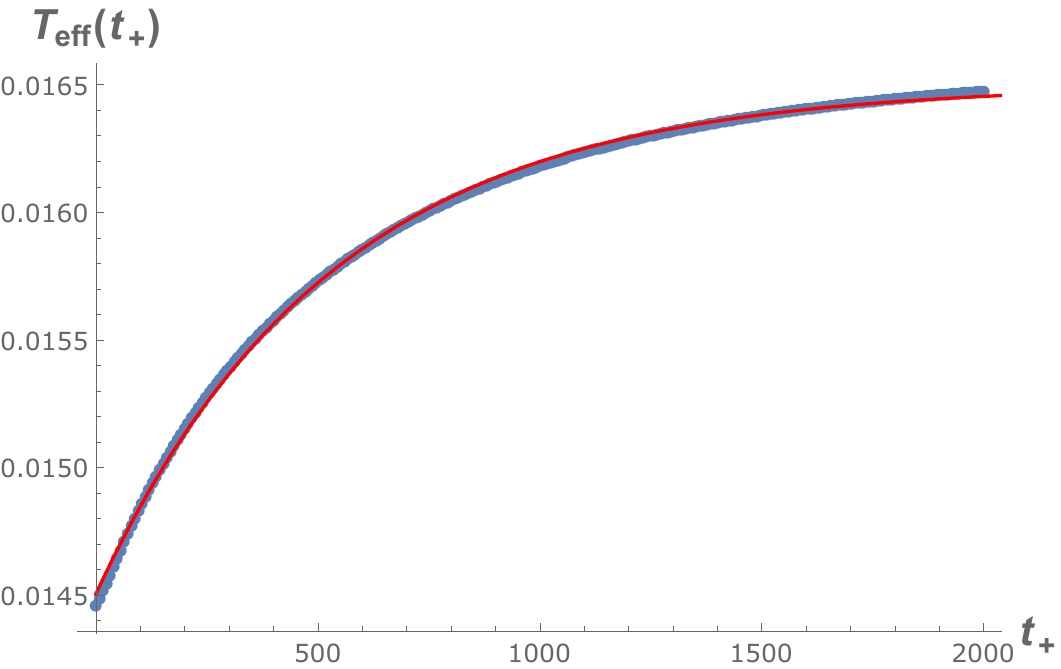}
  \caption{$J_2=0.5$}
  \label{fig:Jt05}
\end{subfigure}%
\begin{subfigure}{.5\textwidth}
  \centering
  \includegraphics[width=.9\linewidth]{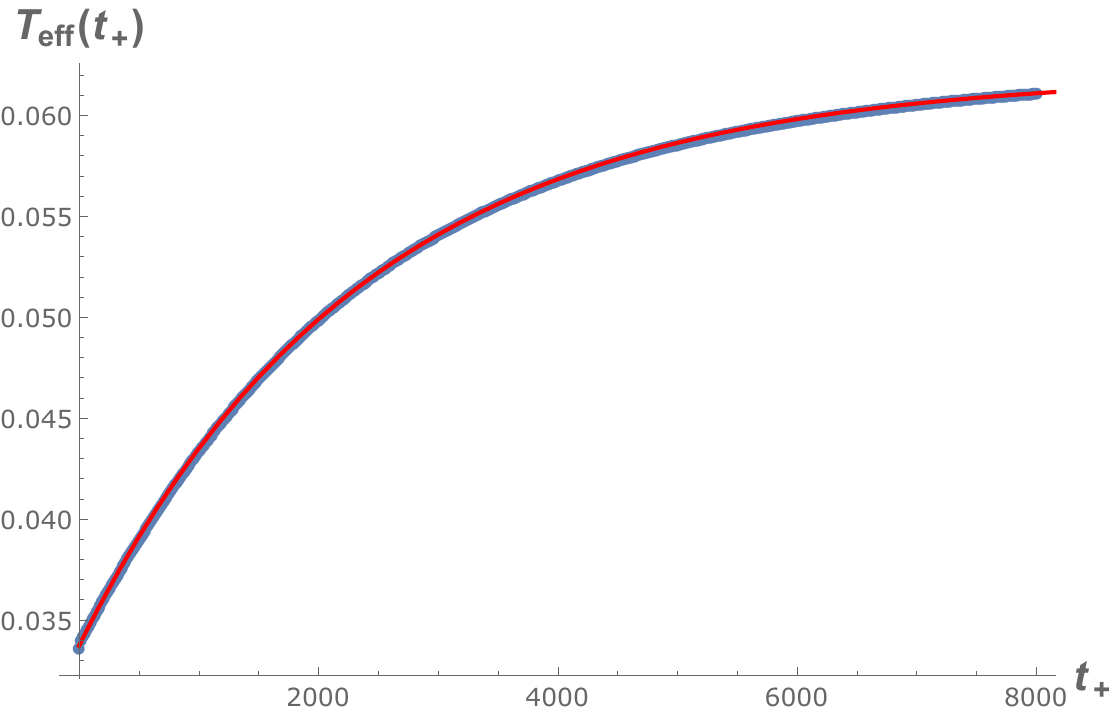}
  \caption{$J_2=2$}
  \label{fig:Jt20}
\end{subfigure}
\caption{\small{Thermalization of $(q=2,4)$ SYK model. (a) The blue dots are the calculated effective temperature for the case of $J_2=0.5, J_4=1$ as a function of $\,t_+$. The initial inverse temperature is $\beta_i=70$. The final inverse temperature is $\beta_f=60.7$. The red curve is an exponential fit with $0.0165-0.0020\times e^{-0.0019 \,t_+}$. (b) The blue dots are the calculated effective temperature for the case of $J_2=2, J_4=1$ as a function of $\,t_+$. The initial inverse temperature is $\beta_i=30$. The final inverse temperature is $\beta_f=17.1$. The red curve is an exponential fit with $0.0620-0.0283\times e^{-0.00042 \,t_+}$.}}
\label{fig:J2_therm}
\end{figure}

Here also we will perform bump quenches. We work with $dt=0.02$. With $J_2=0.5, J_4=1$, the initial state is at inverse temperature $\beta_i=70$. We took the time range $t_1-t_2 \in \{-5000\times dt, 5000\times dt\}$. The quantum quench is performed by turning on $J_6=0.4$ for the time duration $9\times dt$ from time $t=1\times dt$ to $t=9\times dt$. The final inverse temperature is $\beta_f=60.7$. Figure \ref{fig:Jt05} is the plot of the effective temperature as a function of $\,t_+$. With $J_2=2, J_4=1$, the initial state is at inverse temperature $\beta_i=30$. The time range was $\{-3000\times dt,3000 \times dt\}$. The quantum quench is performed by turning on $J_6=0.7$ for the time duration $9\times dt$ from time $t=1\times dt$ to $t=9\times dt$. The final inverse temperature is $\beta_f=17.1$. Figure \ref{fig:Jt20} is the plot of the effective temperature as a function of $\,t_+$. The system thermalizes exponentially fast.

\section{Conclusions}
\label{conclusions}
We show that the liquid state in SYK model with complex fermions thermalizes. The presence of chemical potential suppresses the Lyapunov exponent. The effective temperature equilibrates exponentially fast. Closer examination reveals that the effective temperature is non-monotonic. Without chemical potential, there is a single bump and the effective temperature settles down to its final value. In the presence of the chemical potential, there are damped oscillations during thermalization. The frequency of the oscillations depends on the frequency cut-off used to calculate the effective temperatures.

We also show that the dilute gas phase in SYK model with complex fermions thermalizes instantaneously. The process of instantaneous thermalization suggests that the underlying physics must be different from the well-known chaotic dynamics. We argued that this is because of the long-lived quasi-particle nature of the excitations at very low density in this phase. Moreover, in case of pure $(q=2)$ SYK model, we have also shown that the system equilibrates instantaneously although the final state is not a thermal state.

On the other hand, the $(q=2,4)$ SYK model always thermalizes expoenentially fast. This happens even when the normalized Lyapunov exponent is extremely small $\lambda^*_L\sim 0.003$ for $J_2=2, J_4=1$ at inverse temperature $\beta=17.1$. With these interaction strengths, the spectral statistics of the model is almost completely Poisson statistics. So in conclusion, despite the nature of the spectral statistics, the system is always in a highly chaotic liquid phase without quasi-particle excitation.

\section*{Acknowledgement}
NS thanks Anamitra Mukherjee and Sayantani Bhattacharyya for helpful discussions on this work. NS thanks Krishnendu Sengupta, Ehud Altman and Sumilan Banerjee for communications on different topics related with this work. TS acknowledges financial support from the Department of Atomic Energy, Government of India, for the Regional Centre for Accelerator-based Particle Physics (RECAPP), Harish-Chandra Research Institute.
\bibliography{sykquench} 
\bibliographystyle{JHEP}

\end{document}